\documentclass[nonblindrev]{informs3}

\usepackage{natbib} 
 \bibpunct[, ]{(}{)}{,}{a}{}{,}%
 \def\bibfont{\small}%
\usepackage{multirow}
\usepackage{slashbox}
\usepackage{caption}
\usepackage{subcaption}
\usepackage{multirow}
\usepackage[table]{xcolor}
\usepackage{longtable}
\usepackage{supertabular}

\usepackage{colortbl}

\usepackage{graphicx}
\usepackage{graphics}
\usepackage{Input}
\usepackage{amsmath,amsfonts,latexsym,amssymb}
\usepackage{ifthen}
\usepackage[gen]{eurosym}
\usepackage{colortbl}
\usepackage{wasysym}

\usepackage{todonotes}

\newcommand{\hide}[1]{\ifthenelse{\boolean{includeHidden}}{{\tiny\textbf{HIDDEN:~}#1}}{}}

\newboolean{includeHidden}
\setboolean{includeHidden}{false}

\makeatletter
\def\munderbar#1{\underline{\sbox\tw@{$#1$}\dp\tw@\z@\box\tw@}}
\makeatother

\usepackage{tikz}
\usetikzlibrary{calc,fit}
\usepackage{enumitem}
\makeatletter
\tikzset{%
  remember picture with id/.style={%
    remember picture,
    overlay,
    save picture id=#1,
  },
  save picture id/.code={%
    \edef\pgf@temp{#1}%
    \immediate\write\pgfutil@auxout{%
      \noexpand\savepointas{\pgf@temp}{\pgfpictureid}}%
  },
  if picture id/.code args={#1#2#3}{%
    \@ifundefined{save@pt@#1}{%
      \pgfkeysalso{#3}%
    }{
      \pgfkeysalso{#2}%
    }
  }
}

\def\savepointas#1#2{%
  \expandafter\gdef\csname save@pt@#1\endcsname{#2}%
}

\def\tmk@labeldef#1,#2\@nil{%
  \def\tmk@label{#1}%
  \def\tmk@def{#2}%
}

\tikzdeclarecoordinatesystem{pic}{%
  \pgfutil@in@,{#1}%
  \ifpgfutil@in@%
    \tmk@labeldef#1\@nil
  \else
    \tmk@labeldef#1,(0pt,0pt)\@nil
  \fi
  \@ifundefined{save@pt@\tmk@label}{%
    \tikz@scan@one@point\pgfutil@firstofone\tmk@def
  }{%
  \pgfsys@getposition{\csname save@pt@\tmk@label\endcsname}\save@orig@pic%
  \pgfsys@getposition{\pgfpictureid}\save@this@pic%
  \pgf@process{\pgfpointorigin\save@this@pic}%
  \pgf@xa=\pgf@x
  \pgf@ya=\pgf@y
  \pgf@process{\pgfpointorigin\save@orig@pic}%
  \advance\pgf@x by -\pgf@xa
  \advance\pgf@y by -\pgf@ya
  }%
}

\makeatother

\include{hyphenation}
\usepackage[vlined, ruled]{algorithm2e}

\TheoremsNumberedThrough     

\EquationsNumberedThrough    

\MANUSCRIPTNO{0001}

\begin{document}


\RUNAUTHOR{Bichler, Waldherr}

\RUNTITLE{Combinatorial exchanges with financially constrained buyers}

\TITLE{Competitive Equilibria in Combinatorial Exchanges with Financially Constrained Buyers: \\Computational Hardness and Algorithmic Solutions}

\ARTICLEAUTHORS{%
\AUTHOR{Martin Bichler, Stefan Waldherr}
\AFF{Department of Informatics, Technical University of Munich, Germany}

} 






\ABSTRACT{%
Advances in computational optimization allow for the organization of large combinatorial markets. 
We aim for allocations and competitive equilibrium prices, i.e. outcomes that are in the core. 
The research is motivated by the design of environmental markets, but similar problems appear in energy and logistics markets or in the allocation of airport time slots. Budget constraints are an important concern in many of these markets. 
While the allocation problem in combinatorial exchanges is already $NP$-hard with payoff-maximizing bidders, we find that the allocation and pricing problem becomes even $\Sigma_2^p$-hard if buyers are financially constrained. 
We introduce mixed integer bilevel linear programs (MIBLP) to compute core prices, and propose pricing functions based on the least core if the core is empty. We also discuss restricted but simpler cases and effective computational techniques for the problem. In numerical experiments we show that in spite of the computational hardness of these problems, we can hope to solve practical problem sizes, in particular if we restrict the size of the coalitions considered in the core computations.  
}


\maketitle
{\begin{center} Version \today \end{center}}

\section{Introduction}

Policy-makers but also managers in the private sector have increasingly adopted market mechanisms as a tool to allocate or reallocate scarce resources. These markets need designing and they typically have the following characteristics: (1) there are multiple heterogeneous and indivisible objects, and (2) market participants have complex preferences (objects might be substitutes or complements), allocation constraints and buyers often have budget constraints. We refer to such markets where the underlying allocation problem is a non-convex optimization problem as non-convex markets.

Single-sided combinatorial auctions are one example, and there has been substantial interest in such mechanisms in the past decade. Combinatorial auctions allow bidders to specify package bids, i.e. a price is defined for a subset of the items for auction. The price is only valid for the entire package and the package is indivisible. Moreover, the bidder might have no value for any of these items if he cannot obtain the whole package. For example, in a combinatorial auction, a bidder might want to buy 10 units of item $x$ and 20 units of item $y$ for a package price of \$100, which might be more than what he is willing to pay for the items $x$ and $y$ individually. 
Combinatorial auctions have found widespread application for the sale of spectrum licences \citep{Bichler2017Handbook}, in truck-load transportation \citep{Caplice2006}, for tendering bus routes \citep{Cantillon2006}, or in industrial procurement \citep{Bichler2006}. In addition to the impact on practice, research on combinatorial auctions has resulted in a significant extension of the theoretical literature on markets.  

Unfortunately, the theory of combinatorial exchanges with multiple buyers and sellers is much less well developed. A combinatorial exchange allows participants on both sides to submit bids on packages of indivisible objects. Such types of markets have significant potential for the private and the public sector. Examples include the allocation of airport time-slots \citep{Pellegrini2012, ball2017use}, catch-shares for fishery management \citep{Innes14}, day-ahead energy markets \citep{martin2014strict}, emission-trading \citep{Sadegheih2011}, port capacity \citep{Strandenes2005},  supply chain co-ordination \citep{Fan2003,Guo2012, Walsh2000}, transportation \citep{Schwind2009}, and native vegetation offsets \citep{Nemes2008}. All these examples require bids on packages and they typically involve multiple buyers and sellers. Let us briefly highlight two of these examples in more detail to motivate this research.  

\subsection{Exchanges for Fishery Access Rights}

We will first introduce a concrete example of a combinatorial exchange for fishery access rights that has started this research. 
Catch share systems have been shown to be effective tools to combat overfishing \citep{birkenbach2017catch}, one of the prime environmental concerns these days. Yet, the allocation of catch shares has always been a challenging policy problem \citep{rosenberg2017marine, lynham2014have}. 
There is an active discussion about market-based solutions for the allocation and re-allocation of catch shares \citep{marszalec2017auctions, kominers2017invitation}.  Unfortunately, the design of such markets is challenging as complex constraints need to be considered in practice. 

The recent share trading market in New South Wales (NSW) facilitated the reallocation of catch shares and it is an example for a large-scale combinatorial exchange.\footnote{\url{https://www.dpi.nsw.gov.au/fishing/commercial/reform/historical-docs/adjustment-subsidy-program}} 
The market consisted of several regions, and in each region a number of different types of fishing access rights (aka. share classes) exist. Shares describe the permission to catch a certain type and quantity of fish in a particular region, and each of the 100 share classes consists of a number of shares describing units of effort, such as the number of hooks allowed for line-fishing, the number of nets for net-fishing, etc. 
These shares were distributed evenly among fishers already in the 90s, but they were only effectuated in 2017. After this decision, there was a large percentage of the around 1000 fishers who caught much less than what was allowed, while others needed additional shares. Bilateral bargaining among fishers turned out to be difficult, due to the large number of geographically-dispersed fishers and because of synergistic values among the shares. For example, fishers who wanted to exit the market did not want to be left with subsets of their endowment. They typically wanted to sell all their shares or nothing. Also, buyers wanted to specify specific minimum and maximum quantities of shares to make fishing in a particular region viable. Therefore, package bidding was imperative to facilitate the exchange. These requirements led to a large-scale combinatorial exchange with more than 600 fishers participating and exchanging catch shares.

\subsection{Exchanges for Airport Time Slots}

Combinatorial auctions and exchanges have long been proposed as a market mechanisms for the allocation of airport time slots \citep{rassenti1982combinatorial, Castelli2011, Pellegrini2012, ball2017use}. Package bids are essential in this domain: An takeoff slot at a flight originating airport is only valuable with a landing slot at the flight destination airport. 

Although markets for airport time slots have not yet been implemented, there is an increasing pressure to do so. The growth in air traffic in the past decades has made airport capacity a very scarce resource. Actually, the lack of airport capacity is nowadays a major constraint for the development of air traffic because building of new runways is strongly limited due to cost, environmental impact, land availability, or political reasons. \cite{ball2017use} argue that nowadays there is a strong case for the use of market mechanisms, but these mechanisms need to consider all relevant constraints to be accepted in practice.


\subsection{Budget Constraints}

In both of these markets, one can expect that at least some of the buyers are financially constrained. For example, we experienced that buyers had significant net present values for shares in the exchange for long-term fishery access rights, but they often were financially constrained. In theory, one might assume efficient financial markets such that fishers could raise enough funding, but this is unrealistic to assume in practice. Similarly, one can assume that small airlines have budget constraints such that they are not able to purchase landing rights for which they would have a high net present value. 
Ignoring such financial constraints leads to depressed bidding as buyers can only express budget-capped valuations if they want to avoid a loss. Indeed, some markets elicit both, valuations and budgets that must not be exceeded \citep{nisan2009google}, in order to allow bidders to adequately express their preferences and constraints. 

The importance of budget constraints in practice has led to significant research in mechanism design. Unfortunately, it was shown that we cannot hope for incentive-compatible mechanisms in the presence of budget constraints in multi-object auctions with private budget constraints \citep{Dobzinski2008}. Incentive-compatibility might be too much to ask for and less of a concern in large markets such as the ones introduced above where participants often have very little information about the preferences of others and strategic manipulation is challenging to say the least. However, even if strategic manipulation is less of a concern, a designer might be interested in stable outcomes, where no participant would want to deviate. 

Stability is often seen as a first-order goal in market design \citep{roth2002economist}. The core is the most well-known notion of stability in game theory and it is natural to ask for core-stable outcomes of a market. Such an outcome requires that there cannot be any coalition of participants to have incentives to deviate. 
It was shown that core-stability coincides with the notion of competitive equilibria in markets with payoff-maximizing bidders and that the core might be empty in a combinatorial exchange \citep{Bikhchandani2002}. Unfortunately, it turns out that the presence of budget constraints makes the computation of core-stable outcomes a much harder computational problem. The problem actually becomes $\Sigma_2^p$-hard, which is interesting as not many practical problems fall into this complexity class. In this paper, we analyze the problem theoretically and suggest algorithmic approaches to compute stable outcomes. Experiments should illustrate that such approaches allow us to solve surprisingly large problem sizes in practice. 

\subsection{Related Literature}

Let us first provide a brief survey of the relevant literature in computer science, economics, and operations research. Mechanism design was successfully applied for the design of one-sided auctions. The traditional mechanism design literature assumes preferences where bidders have independent and private valuations and quasilinear utility functions, i.e. participants are pure payoff-maximisers. The literature imposes budget-balance, individual rationality, incentive-compatibility, and efficiency (i.e. welfare maximization) as primary design goals and models games with incomplete information. The fact that the Vickrey-Clarke-Groves (VCG) mechanism is dominant-strategy incentive compatible is remarkable given basic impossibility results in social choice theory \citep{Gibbard1973,Satterthwaite1975}. Unfortunately, incentive-compatibility is conflicting with other design desiderata. \cite{Myerson1983} have already proven that there is no market mechanism which allows the achieving of all four design desiderata mentioned above. For single-sided auctions, the Vickrey-Clarke-Groves (VCG) mechanism satisfies the four design desiderata but, in environments with multiple buyers and sellers, the VCG mechanism is not budget-balanced and the auctioneer might make a substantial loss. Budget-balance is almost always a hard constraint.

A few authors have proposed pricing rules for combinatorial exchanges. \cite{Parkes01} suggests a threshold scheme, which gives surplus to agents with payments further than a certain threshold value from their Vickrey payments. \cite{Lubin08} suggest a tree-based bidding language as well as simple linear prices in an iterative combinatorial exchange design. We add to this literature, but aim for payments that yield core stability in our paper.

Note that payoff-maximization in the form of a quasi-linear utility function is also an important assumption for the VCG mechanism to be strategy-proof, i.e. dominant-strategy incentive-compatible. If bidders have private budget constraints, as we assume in this article, quasi-linearity is violated and  incentive-compatible auction mechanisms do not exist anymore for multi-object markets \citep{Dobzinski2008, Fadaei2017a}. 
There has been recent progress on approximation mechanisms for two-sided combinatorial auctions \citep{Colini-Baldeschi16c}, but this work does not focus on budget constraints or the core as a design goal. Overall, the mechanism design literature shows that those environments which allow for incentive-compatible mechanisms are quite limited. 

Strategic manipulation is less of a concern in large markets such as the market for fishery access rights. Already, \cite{Roberts1976} showed that in large markets the ability of an individual player to influence the market is minimal, so agents should behave as price-taking agents. In addition, in a combinatorial exchange bidders can submit bids on an exponential set of packages and neither the type nor the number of bidders participating is known in advance, which makes strategic manipulation much more challenging. 

The complete-information analysis has been the standard approach in general equilibrium theory, and it is natural to first understand efficiency and payments of a market in a complete-information model with price-takers. Stability in the form of competitive equilibria and efficiency are the central design goals in general equilibrium theory. The celebrated Arrow--Debreu model suggests that, under certain assumptions such as divisible objects, convex preferences, and demand independence there must be a set of anonymous and linear (i.e. item) prices such that aggregate supplies will equal aggregate demands for every commodity in the economy \citep{Arrow1954}. 

Unfortunately, also more recent literature on general equilibrium theory is restricted to markets with divisible goods and the results do not carry over to combinatorial markets \citep{oneill2005efficient}. However, stability and efficiency are important design desiderata for combinatorial exchanges. The core is probably the most important solution concept for coalitional games \citep{aumann2006war}.
There has been limited literature on the equivalence of the core and competitive equilibria in combinatorial exchanges \citep{Bikhchandani2002,bichler2017core}. 
This literature is rooted in linear programming and duality theory. However, it assumes bidders with pure quasi-linear utility functions and does not allow for budget constraints. For the design of real-world markets, it is important to consider such financial constraints as well, which is what we do in this paper. 
Unfortunately, budget constraints have substantial impact on allocation and payment functions of a mechanism as we show, because these two problems are not separable anymore. On a more general level, \cite{roughgarden2015prices} highlight the tight connection between pricing, algorithms, and optimization and our work contributes to this line of research.

\subsection{Contributions}

We aim for welfare maximization subject to budget constraints and core-constraints in large non-convex markets. The allocation problem in a combinatorial exchange is already $NP$-hard. Not many practical problems fall in higher classes of the polynomial hierarchy. However, our computational complexity analysis actually yields that in the presence of budget-constrained buyers the allocation and pricing problem becomes $\Sigma_2^p$-hard. This is important to show formally, and requires an elaborate reduction from the canonical $\Sigma_2^p$-hard problem QSAT$_2$.

Mixed integer bilevel linear programs (MIBLPs) allow us to model such problems. While bilevel programming has been a topic in the operations research literature for many years, algorithms to solve MIBLPs have only seen progress in the recent years. This provides us with an opportunity to further develop algorithmic approaches for MIBLPs in a practically relevant domain. As the core can sometimes be empty, we also propose pricing functions based on the least core.
We introduce effective cuts to reduce the number of coalitions that need to be considered, as well as decomposition approaches that allow us to reduce the size of the problem considerably in practice. In addition, we also show restricted cases that reduce the computational complexity to the class of NP-hard optimization problems. 

In numerical experiments we show that in spite of the computational hardness of these problems, we can solve practical problem sizes, in particular if we restrict the size of the coalitions considered in the core computations. Such restrictions keep the problem tractable for realistic problem sizes and might provide a sufficient level of stability for practical applications.  Replicability of experimental results is an important concern and, therefore, we use the CATS test suite \citep{leyton2000towards}, which provides a widely used instance generator for the airport time-slot allocation problem discussed earlier.

\section{Model and Preliminaries}
\label{sec:model}

We first introduce a model without budget constraints based on \cite{Bikhchandani2002} and \cite{bichler2017core}. The papers show equivalence of the core and competitive equilibria in a combinatorial exchange by drawing on specific linear programming formulations. This will be our starting point for the analysis of budget constrained buyers.

There is a finite set of bidders $N$, consisting of buyers $i \in I$ and sellers $j \in J$ with $I \cup J = N$ and $I \cap J = \emptyset$, as well as a finite set of indivisible objects or items, $K$. Each buyer $i \in I$ has a non-negative valuation for each set of objects $S \subseteq K$ denoted $v_i(S) \in \mathbb{R}_{\geq 0}$ with $v_i(\emptyset)=0$. 

Sellers also have valuations or reservation prices for packages $Z \subseteq K$ with $v_i(Z) \in \mathbb{R}_{\geq 0}$. Buyers and sellers have free disposal. Every package is priced and each buyer $i \in I$ pays the price $p_i(S)$ for the bundle $S$ he receives, and each seller $j \in J$ receives the payment $p_j(Z)$ for the bundle $Z$ he supplies. The vectors $P_i=(p_i(S))_{i,S}$ and $P_j=(p_j(Z))_{j,Z}$ describe the non-linear prices of buyers and sellers. 
In our initial analysis the preferences are quasi-linear, i.e. the payoff of the buyer is $\pi_i = v_{i}(S) - p_i(S)$ and that of the seller is $\pi_j = p_j(Z) - v_{j}(Z)$. Later we will add budget constraints.

The problem of finding an efficient assignment maximizing gains from trade among buyers and sellers can be formulated as a linear program as follows: We use binary variables $x_i(S)$ to describe whether package $S$ is assigned to bidder $i$ and $y_j(Z)$ to describe whether package $Z$ is supplied by seller $j$. The vectors $X=(x_i(S))_{i,S}$ and $Y=(y_j(Z))_{j,Z}$ describe the allocations of buyers and sellers. The model enumerates all possible allocations similar to the single-seller model in \cite{de2007ascending}. The set of all possible object assignments is denoted as $\Gamma$, a specific assignment as $(X,Y) \in \Gamma$. For each possible allocation, we have a binary variable $\delta_{X,Y}$, which is one if an allocation is selected and zero otherwise. The model allows for a very natural interpretation of the dual variable as prices. The dual variables of \textbf{P} are written in brackets. 

\begin{small}
\[\begin{array}{rcll@{\hspace{2em}}l}
w_P=\max & \sum_{i\in I}\sum_{S \subseteq K}v_{i}(S)x_{i}(S) - \sum_{j\in J}\sum_{Z \subseteq K}v_{j}(Z)y_{j}(Z)&\label{eq:wdp} \tag{\textbf{P}} & \\
\textrm{s.t.} & x_{i}(S)-\sum_{x: x_{i}=S}\delta_{X,Y}=0 & \forall i\in I,\forall S\subseteq K & (p_{i}(S))\\
 & -y_{j}(Z)+\sum_{y: y_{j}=Z}\delta_{X,Y}=0 & \forall j\in J,\forall Z\subseteq K & (p_{j}(Z))\\
 & \sum_{S\subseteq K}x_{i}(S)\leq 1 & \forall i \in I & (\pi_{i})\\
 & \sum_{Z\subseteq K}y_{j}(Z)\leq 1 & \forall j \in J & (\pi_{j})\\
 & \sum_{(X,Y)\in\Gamma}\delta_{X,Y} = 1 & & ( \pi_a)\\
 & 0\leq x_{i}(S) & \forall S\subseteq K,\forall i\in I&\\
 & 0\leq y_{j}(Z) & \forall S\subseteq K,\forall j\in J&\\
 & 0\leq\delta_{X,Y} & \forall (X,Y)\in\Gamma&\\
 \end{array} 
\]
\end{small}

The formulation \textbf{P} introduces a variable $\delta_{(X,Y)}$ for each possible allocation, making the linear program large but integral. An LP solver selects one vertex with $\delta_{(X,Y)}=1$, such that we always get integer allocations $x_i(S)$ and $y_j(Z)$ of \textbf{P} \citep{bichler2017core}. At least one of these allocations maximizes the gains from trade, i.e. welfare in the economy. 

With a single seller, the problem is equivalent to the winner determination problem in combinatorial auctions, which is already known to be $NP$-hard \citep{lehmann2006winner}.\footnote{Note that even though the problem is $NP$-hard, there are algorithms that run in polynomial time in the size of the input if the number of bids is very large compared to the number of items \citep{lehmann2006winner}.} Note that there are more effective formulations as binary program, that we will use in Section \ref{sec:miblp}, where we discuss a bilevel program to compute core payments in the presence of financially constrained bidders. However, model \textbf{P} nicely shows core payments can be computed without budget constraints.

The core-prices resulting from the dual variables of \textbf{P} are non-linear and personalized. It is straightforward to show that linear prices are impossible if bidders have complementary valuations \citep{Kelso82}. We can now formulate the dual \textbf{D} of \textbf{P}.

\[\begin{array}{rcll@{\hspace{2em}}l}
\min & \sum_{i \in I} \pi_i + \sum_{j \in J} \pi_j + \pi_a & \label{eq:wdd} \tag{\textbf{D}}& \\
\textrm{s.t.} & \pi_i \geq v_i(S) - p_i(S) & \forall i \in I, \forall S \subseteq K & (x_i(S))\\
 & \pi_j \geq p_j(Z) - v_j(Z) & \forall j \in J, \forall Z \subseteq K & (y_j(S))\\
 & \sum_{y_{j}(Z) \in Y} p_j(Z) - \sum_{x_{i}(S) \in X} p_{i}(S)  + \pi_a \geq 0 & \forall (X,Y) \in \Gamma & (\delta_{X,Y})\\
 & \pi_{i}, \pi_j, p_i(S), p_j(Z) \geq 0 & \forall S,Z \subseteq K, \\
 & &\forall i\in I, \forall j \in J &\\
 & \pi_a \in \mathbb{R} \\
 \end{array} 
\]

\cite{bichler2017core} show that if $\pi_a = 0$, an optimal solution of \textbf{D} lies in the \textit{core} of the auction. 

\begin{definition}
Let $\Pi_i=(\pi_i) \in \mathbb{R}^{|I|}_{\geq 0}$ and $\Pi_j=(\pi_j) \in \mathbb{R}^{|J|}_{\geq 0}$ be the payoff vectors of the buyers and sellers in the auction. Then $(\Pi_i,\Pi_j)$ is in the \textit{core} of the game $\mathcal{E}$, denoted $(\Pi_i,\Pi_j) \in \textrm{core}(\mathcal{E}$), if
\[\begin{array}{rcll@{\hspace{2em}}l}
 & \sum_{i\in I} \pi_i + \sum_{j\in J} \pi_j = V(N) &  & \textrm{core efficiency}\\
 & \sum_{i\in \mathcal{C}} \pi_i + \sum_{j\in \mathcal{C}} \pi_j \geq V(C) & \forall C \subset N=I \cup J &\textrm{core rationality}
 \end{array} \]
\end{definition}

More importantly, if the core of the auction is non-empty, then the set of optimal solutions of \textbf{D} with $\pi_a = 0$ coincides with the core. Since the dual can be solved via a linear program, once can determine in polynomial-time, whether the core is empty or not.

The presence of budget constraints $B_i$ of the buyers $i \in I$ violates quasi-linearity and, for the markets that we analyze, the budget constraints of buyers cannot be ignored. Quasi-linear utility functions describe an environment with transferable utility. If we have budget constraints, only parts of the utility of the buyer up to the budget is indeed transferable.


\hide{ 

	\subsection{Core and pricing equilibrium for CAs and CEx}
	
The complexity of problems associated with solution concepts for cooperative games has long been a subject in the literature. In the following, we summarize the results for cooperative games in general as well as for combinatorial auctions and exchanges. 

We first concentrate on the case with transferable utility. \cite{deng1994complexity} were the first to discuss the complexity of core computations for graph games. In these games, each player corresponds to an edge in an undirected graph and the value of each coalition is calculated by the combined weight of edges between its members. They show that for these types of games, determining whether the core is non-empty and showing for an imputation whether it is in the core are $co-NP$-hard problems if the edges are allowed to be negative, but can be solved in polynomial time if all edges are non-negative. These results carry over for hypergraphs if the number of edges is polynomial in the number of players. \cite{ieong2005marginal} introduce marginal contribution networks as another compact representation for coalition games and proof the same complexity results for these games. \cite{greco2011complexity} generalize these results. They distinguish between $P$-representations where the worth of coalitions can be computed in polynomial time (as in graph games and marginal contribution networks) and $NP^{OPT}$-representations of a game, where the worth function is an $NP$-hard optimization problem. Moreover, they introduce the notion of $wNP^{OPT}$-representations in which the worth of $v(N)$ of the whole set N is explicitly provided or can be calculated easily. They show that deciding whether the core is non-empty and to determine whether an imputation is in the core are $coNP$-complete problems for $P$- as well as $wNP^{OPT}$ representations, while for $NP^{OPT}$ deciding non-emptiness of the core is $\Delta_2^P$-complete and checking core-membership of an imputation is $D^P$-complete.

For games with non-transferable utility, \cite{malizia2007infeasibility} prove that the complexity of deciding emptiness of the core is $\Sigma_2^p$-complete even for games which are given in a generalized form of the marginal contribution network representation by \cite{ ieong2005marginal}.

\cite{deng1999algorithmic} discuss combinatorial optimization games where the game value is calculated by an integer linear program. They show that the core for a these combinatorial optimization games is nonempty if and only if the corresponding LP relaxation has an integer optimal solution. For several combinatorial optimization games it shown that testing of non-emptiness of the core can be done in polynomial time, even for $NP$-complete problems (e.g. max independent set). This is related to the result by \cite{bikhchandani2002package} who discuss competitive equilibria and core assignments in combinatorial auctions and exchanges. Therein, interconnections among the linear programming formulation, Walrasian equilibrium, and the core are established. In the single seller (auction) version, a necessary and sufficient condition is given for the Vickrey payoff point to be implementable by a pricing equilibrium. Higher-order price equilibria are also core outcomes as Walrasian equilibria for first-order price equilibria. With multiple sellers, there might not be competitive equilibrium as the solution might not be integral. \cite{ bikhchandani2002package} show that a competitive equilibrium in the package assignment model exists if and only if the associated linear program is integral, in which case the dual solutions form the set of pricing equilibria. \cite{bichler2017core} extend the models by \cite{bikhchandani2002package} and provide a model for combinatorial exchanges, where the core is equivalent to the set of competitive equilibrium prices, if the core is non-empty.

In combinatorial auctions, the core is always non-empty since the efficient allocation and charges each bidder his full bid amount is in the core. Moreover, as follows from above, if the efficient allocation (i.e. the worth of the grand coalition) is given, checking for an allocation whether it is in the core is a $co-NP$-complete problem. In comparison, based on \cite{bikhchandani2002package}, the core might be empty in combinatorial exchanges with more than one seller. 

\begin{itemize}
		\item How hard is it to compute core imputations for CEs? \textbf{OPEN}
		\item Checking whether the core is non-empty or not in itself is done in polynomial time in the number of allowed coalitions because it is formulated as a linear programming problem (see below). However, \cite{Conitzer06} pointed out that it requires many constraints for all the sub-coalitions and \textbf{the size of the representation (input) is exponential in the number of agents}. Any algorithm for computing the core payoff vector requires time exponential as long as it reads all the input. 
		\item If a coalition is prohibited (e.g., due to allocation constraints) or if its value is not explicitly specified, the algorithm may need to compute every value of such coalitions. Computing a value of a coalition is not necessarily straightforward because the agents must solve a complex collaborative planning problem. 
		\item In general, computing an optimal coalition structure is known to be NP-hard and checking whether there exists a core for the (optimal) coalition structure or not is NP-complete unless its value is explicitly provided. QUESTION: \textbf{Is the computation in NP, even though one has to compute $V(C)$ for an set of coalitions exponential in the number of agents?}
		\item Deng et al. claim that finding an imputation in the core of the max. independent set game is in P. They also say that the core is non-empty iff the LPR is integral. \textit{This is violated by examples of single-seller CAs which have a core, but are not integral.}
\end{itemize}

\section{Model}

We consider a finite set of bidders $N$, consisting of buyers $i \in I$ and sellers $j \in J$ with $I \cup J = N$ and $I \cap J = \emptyset$, as well as a finite set of indivisible objects or items, $K$. Each buyer $i \in I$ has a non-negative valuation for each set of objects $S \subseteq K$ denoted $v_i(S) \in \mathbb{R}_{\geq 0}$ with $v_i(\emptyset)=0$. Sellers also have valuations or reservation prices for packages $Z \subseteq K$ with $v_i(Z) \in \mathbb{R}_{\geq 0}$. Buyers and sellers have free disposal. We use binary variables $x_i(S)$ to describe whether package $S$ is assigned to bidder $i$ and $y_j(Z)$ to describe whether package $Z$ is supplied by seller $j$.

Every package is priced and each buyer $i \in I$ pays the price $p_i(S)$ for the bundle $S$ he receives, and each seller $j \in J$ receives the payment $p_j(Z)$ for the bundle $Z$ he supplies. For each buyer $i$ there may exist an exogenous budget constraint $B_i$ which restricts the payment that $i$ is able to make. In The utility function for each buyer $i \in I$ is given by
\[
	u_i(S) = \begin{cases} v_i(S) - p_i(S) &\textrm{, if } p_i(S) \leq B_i \\
													- \infty &\textrm{, else}
						\end{cases}
\]
which is not quasi-linear.

The combinatorial exchange with budget constraints can be interpreted as a special case of a coalitional game with non-transferable utility (NTU-game), i.e. a pair $(N,V)$ where $N$ is a finite set, the set of players, and, for every coalition $S \subseteq N$, $V(S)$ is a subset of $\mathbb{R}^N$ satisfying the following four conditions:
\begin{enumerate}
	\item If $S \ne \emptyset$, then $V(S)$ is non-empty and closed, and $V(\emptyset)=\emptyset$.
	\item For every $i \in N$ there is a $V_i \in \mathbb{R}$ such that for all $x \in \mathbb{R}^N: x \in V(\{i\})$ if and only if $x_i \leq V_i$
	\item If $x \in V(S)$ and $y \in \mathbb{R}^N$ with $y_i \leq x_i$ for all $i \in S$ then $y \in V(S)$.
	\item $\{x \in V(N):x_i \geq V_i\}$ is a compact set.
\end{enumerate}

The core $\mathcal{C}(N,V)$ of a game $(N, V)$ is the set of payoff vectors $x \in V$ that can not be improved upon by any coalition $C$, i.e. there is no $y \in V(C)$ such that $y_i > x_i$ for all $i \in C$.

} 
\section{Budget Constraints}

In what follows, we aim for welfare maximization subject to budget constraints and core-constraints in large non-convex markets. In other words, we want to compute allocations that maximize welfare, but are stable considering valuations and budgets of bidders. 

\subsection{Complexity Analysis}

First, observe that core constraints might restrict the welfare gains in a combinatorial exchange as the following example shows.

\begin{example}
Suppose there are two buyers, $b_1$ and $b_2$, having a value of \$10 and \$9 for a good, resp. In addition, buyer $b_1$ has a budget constraint of \$1 and cannot spend more money. There are also sellers S1 and S2 with reserve prices of \$0 and \$4. The welfare-maximising allocation is to match $b_1$ and S1 at a price of \$1, and $b_2$ and S2 at a price somewhere between \$9 and \$4, which yields \$15 gains from trade. However, this efficient allocation is not stable, because S1 could approach $b_2$ and they could agree to deviate at a price of less than \$4 and more than \$1, which is profitable for both of them. Matching buyer S1 to $b_2$ is stable, but the gains from trade are only \$9, as compared to the welfare maximizing allocation with gains from trade of \$15.
\end{example}

The example simply illustrates that budget constraints can reduce the gains from trade. Providing valuations and budget constraints on a market is not unusual. For example, in Google's auction for TV ads buyers provided both \citep{nisan2009google}.

From \cite{Bikhchandani2002} we know that the core of a combinatorial exchange can be empty even without budget constraints. But even if the core is non-empty without budget constraints, it can be empty if such constraints are added. 

\begin{theorem}\label{prop:emptycore}
A budget-constrained combinatorial exchange instance may have an empty core, even if the core is non-empty when budgets are ignored.
\end{theorem}

\proof{Proof:}
Consider a case with two sellers, $S_1$ offering item $A$ and $S_2$ offering item $B$ and two buyers with the following valuations and budgets:\\

\begin{table}[ht]
\centering
\caption{Example}
\label{ex:core}
\begin{tabular}{|l|l|l|l|l|}
\hline
& \{A\} & \{B\} & \{A,B\} & Budget\\
\hline
Buyer $b_1$ & 0 & 0 & 10 & 3\\
Buyer $b_2$ & 4 & 4 & 4 & 2\\
\hline
\end{tabular}
\end{table}

Without budget constraints, buyer $b_1$ can pay a price of 4 for each of the items, resulting in a core outcome. However, if we consider the budget constraints, there is no core outcome. Suppose buyer $b_1$ obtains $\{A,B\}$ for a combined price of at most 3. Then there is at least one seller with a payoff lower than $2$ and buyer $b_2$ and this sellers can form a coalition in which both are better off. Similarly, suppose $b_2$ obtains one of his desired items at one of the sellers while $b_1$ does not obtain any items. Since the combined payoff of the sellers is at most $2$, $b_1$ and the sellers can form a coalition where all three are better off. 
\endproof

The proof of Theorem \ref{prop:emptycore} illustrates that it can be quite intricate to detect whether the core of a combinatorial exchange is empty or not.
A combinatorial exchange with budget constraints on the buyers' side can be seen as a game with partially transferable utility. The problem has a specific structure and therefore is important to understand its computational complexity.
 
\begin{theorem}
Computing a welfare-maximizing core outcome in a combinatorial exchange with budget constraint is $\Sigma_2^p$-hard
\end{theorem}

Proof techniques for this complexity class are much less developed than those for lower levels in the polynomial hierarchy. The proof (see Appendix) reduces from QSAT$_2$ and requires an elaborate construction. A reduction from more abstract problems such as min max clique, which are known to be $\Sigma_2^p$-hard, appears simpler at first sight, but has shown not to be straightforward. 

An interesting question is whether there are some sufficient conditions for the core of the combinatorial exchange that are simple to check. 
For transferable utility games the Bondareva-Shapley theorem describes balancedness as a \textit{necessary and sufficient} condition for the core of a cooperative game with transferable utility (TU) to be non-empty (see \cite{bondareva1963some} and \cite{shapley1967balanced}). Balancedness is a rather obscure property, but it can be checked with linear programming in transferable utility games, and linear programming was also used to decide whether the core is empty in combinatorial exchanges without budget constraints \citep{bichler2017core}.

A combinatorial exchange with budget constraints is closer to a game with non-transferable utility (NTU). The main result for NTU-games is that balanced games have a nonempty core, but the converse is not true \citep{scarf1967core}. So, balancedness is a requirement that is sufficient for TU and NTU games to have a non-empty core. Scarf's algorithm is a central result to compute whether the core of an NTU game is empty. Unfortunately, this algorithm is already PPAD-complete \citep{kintali2008complexity}. Moreover, the algorithm requires a matrix as an input that has a column with the payoff of each player for every coalition. In a combinatorial exchange with budget constraints, every coalition can have multiple allocations with different payoffs. Moreover, we have partially transferable utility (up to the budget constraint), and the payoff vectors for each coalition are not unique for a coalition as is the case for a pure NTU game, but depend on the prices. This renders Scarf's algorithm not applicable.

\subsection{Computing Core Payments via MIBLPs}
\label{sec:miblp}

Next, we show that mixed integer bilevel linear programs (MIBLP) provide an adequate mathematical abstraction to model the problem. This is a field of mathematical optimization that is notoriously hard, but recent algorithmic advances suggest that such problems can be solved in practice \citep{zeng2014solving, fischetti2017new, tahernejadbranch}. 

The MIBLP we suggest finds core allocations in combinatorial exchanges with budget constraints. If the core is not empty, the solution of the bilevel program consists of a core allocation with maximum welfare. Additionally, we obtain prices and payments for buyers and sellers. If the core is empty, the MIBLP is infeasible. 



The objective of of the MIBLP is to find prices $P=\{P_i,P_j\}$, such that the corresponding allocation $(X,Y)$ is in the core and there is no core allocation with a higher welfare.  We maximize gains from trade as the standard way to maximize welfare in double auctions. 


In the lower level of the bilevel program, for each coalition of bidders $C \subset I \cup J = N$, the smallest improvement $d^C$ of any member of $C$ when deviating from the grand coalition is calculated. For this, given the set of buyers $I(C)$ and sellers $J(C)$ in the coalition, as well as the set of items $\mathcal{K}(C)$ which is endowed to sellers part of $C$, an allocation of bundles $S,Z \subseteq \mathcal{K}(C)$ is determined. Thus, in the lower level, variables $\chi^C_i(S) \in \left\{0,1\right\}$ and $\gamma^C_j(Z) \in \left\{0,1\right\}$ denote the allocation of bundles $S,Z \subseteq \mathcal{K}(C)$ within coalition $C$ and $p^C_i(S), p^C_j(Z)$ describe the corresponding prices and payments. 

Then, the bilevel program can be written as \eqref{eq:wd}. For convenience, the bilevel program is presented with multiple lower levels, one for each coalition $C \subseteq N$ and with (non-linear) multiplications of variables for the allocations and the corresponding price (e.g. in constraint \eqref{eq:ul_bb}). Both can be easily rewritten as a mixed integer linear bilevel program with a single lower level. 

\begin{tiny}
\begin{subequations}
\begin{align} 
 \max_{x_{i}(S), y_j(Z)} & \sum_{S \subseteq K} \sum_{i \in I} v_{i}(S) x_i(S) - \sum_{j\in J}\sum_{Z \subseteq K}v_{j}(Z)y_{j}(Z) \label{eq:wd} \tag{\textbf{CEx}} \\
  \text{s.t.}  &\sum_{S \subseteq K} p_{i}(S) x_i(S) \leq \min\left\{B_i,\sum_{S \subseteq K} v_{i}(S) x_i(S)\right\}  & \forall i \in I \label{eq:ul_bc} \tag{BC}\\\
	&p_{j}(Z) y_j(Z) \geq v_{j}(Z) y_j(Z)  & \forall Z \subseteq K, \forall j \in J \label{eq:ul_irs} \tag{IRS}\\	
	&\sum_{i \in I} \sum_{S \subseteq K} p_{i}(S) x_i(S) = \sum_{j \in J} \sum_{Z \subseteq K} p_{j}(Z) y_j(Z) \label{eq:ul_bb} \tag{BB}\\
	&\sum_{S: k \in K} \sum_{i \in I} x_i(S) \leq \sum_{Z: k \in K} \sum_{j \in J} y_j(Z) & \forall k \in K \label{eq:ul_supply} \tag{Supply}\\
	&\sum_{S \subseteq K} x_i(S) \leq 1 & \forall i \in I \label{eq:ul_xor-b} \tag{XOR-B}\\
	&\sum_{Z \subseteq K} y_j(Z) \leq 1 & \forall j \in J \label{eq:ul_xor-s} \tag{XOR-S}\\
	&d^C \leq 0 & \forall C \subset I \cup J \label{eq:core} \tag{Core}\\
	&d^C = \max d^C & \forall C \subset I \cup J \label{eq:ll} \tag{\textbf{Lower Level}} \\
	&\textrm{s.t.} \sum_{S \subseteq K(C)} p^C_{i}(S) \chi^C_i(S) \leq \min\left\{B_i,\sum_{S \subseteq K(C)} v_{i}(S) \chi^C_i(S)\right\} & \forall i \in I(C) \label{eq:ll_bc} \tag{BC}\\
	&\hspace{0.7cm}p^C_{j}(Z) \gamma^C_j(Z) \geq v_{j}(Z) \gamma^C_j(Z) & \forall j \in J(C), \forall Z \subseteq K \label{eq:ll_irs} & \tag{IRS}\\	
	&\hspace{0.7cm}\sum_{i \in I(C)}\sum_{S \subseteq K(C)} p^C_{i}(S) \chi^C_i(S) = \sum_{j \in J(C)}\sum_{Z \subseteq K(C)} p^C_{j}(Z) \gamma^C_j(Z) \label{eq:ll_bb} \tag{BB}\\
	&\hspace{0.7cm}\sum_{S: k \in K(C)} \sum_{i \in I} \chi^C_i(S) \leq \sum_{Z \subseteq K(C)} \gamma^C_j(Z) & \forall k \in K(C) \label{eq:ll_supply} \tag{Supply}\\ 
	&\hspace{0.7cm}\sum_{S \subseteq K(C)} \chi^C_i(S) \leq 1 & \forall i \in I(C) \label{ll_xor-b} \tag{XOR-B}\\
	&\hspace{0.7cm}\sum_{Z \subseteq K(C)} \gamma^C_j(Z) \leq 1 & \forall j \in J(C) \label{ll_xor-s} \tag{XOR-S}\\
	&\hspace{0.7cm} d^C \leq \sum_{S \subseteq K(C)} \left(v_i(S) - p^C_i(S)\right)\chi^C_i(S) - \notag \\ 
	&\hspace{0.9cm} \sum_{S \subseteq K} \left(v_i(S) - p_i(S)\right)x_i(S) & \forall i \in I(C) \label{eq:ll_core_b} \tag{Imp-B} \\
	&\hspace{0.7cm} d^C \leq \sum_{Z \subseteq K(C)}  p^C_j(Z) \gamma^C(Z) - \sum_{Z \subseteq K} p_j(Z)y_j(Z) & \forall j \in J(C) \label{eq:ll_core_s} \tag{Imp-S} \\
	&\hspace{0.7cm} \chi^C_i(S) \in \{0,1\} & \forall S \subseteq K(C), i \in I(C) \label{eq:ll_$b_1$} \tag{Binary}\\ 
	&\hspace{0.7cm} \gamma^C_j(Z) \in \{0,1\} & \forall Z \subseteq K(C), j \in J(C) \label{eq:ll_$b_2$} \tag{Binary}\\
	&\hspace{0.7cm} p^C_{i}(S) \in \mathbb{R}^+_0 & \forall S \subseteq K(C), i \in I(C) \label{eq:ll_r} \tag{Real} \\
	&\hspace{0.7cm} p^C_{j}(Z) \in \mathbb{R}^+_0 & \forall Z \subseteq K(C), j \in J(C) \label{eq:ll_r2} \tag{Real} \\
	&\hspace{0.7cm} d^C \in \mathbb{R} & \label{eq:ll_r2} \tag{Real}\\
	& x_{i}(S) \in \{0,1\} & \forall S \subseteq K, i \in I \label{leader_b} \tag{Binary}\\
	& y_{j}(Z) \in \{0,1\} & \forall Z \subseteq K, j \in J \label{leader_$b_2$} \tag{Binary}\\
  & p_{i}(S) \in \mathbb{R}^+_0 & \forall S \subseteq K, i \in I  \label{leader_r} \tag{Real}\\
	& p_{j}(Z) \in \mathbb{R}^+_0 & \forall Z \subseteq K, j \in J  \label{leader_r2} \tag{Real}
\end{align}
\end{subequations}
\end{tiny}

The objective of \eqref{eq:wd} is to maximize gains from trade by determining an assignment of packages and corresponding prices, such that the prices respect the budget constraints and individual rationality of buyers \eqref{eq:ul_bc} and sellers \eqref{eq:ul_irs}, and budget balance \eqref{eq:ul_bb}. Further, only items which are sold can be allocated to buyers \eqref{eq:ul_supply}, each buyer may only obtain at most one package \eqref{eq:ul_xor-b} and each seller may only sell at most one package consisting of the items endowed to him \eqref{eq:ul_xor-s}. The prices have to be set in such a way, that no coalition can benefit from deviating.  For each coalition $C$, an assignment $\chi^C,\gamma^C$ with payments $P^C$ is determined in the lower level. Similar to the upper level, these assignments have to respect budget constraints \eqref{eq:ll_bc}, budget balance \eqref{eq:ll_bb} and the supply constraint \eqref{eq:ll_supply}. Particularly, only items offered by sellers part of the coalition may be allocated. Constraints \eqref{eq:ll_core_b} and \eqref{eq:ll_core_s} denote the improvements for each individual buyer and seller when participating in this coalition in comparison to the grand coalition. The objective of the lower level is to maximize the minimum improvement. For the allocation $(X,Y)$ and the corresponding payments to be in the core, this improvement must not be positive for any coalition \eqref{eq:core}.

Instead of multiple lower levels, we can rewrite the lower level optimization problem in vector notation as
\[
d = \max \sum_{C \subset I \cup J} d^C
\]
 and calculate the allocations for each coalition simultaneously in the lower level since variables from different coalitions are independent from each other. Products of continuous price variables $p$ and binary variables $x,y,\chi,\gamma$ can be linearized by introducing auxiliary variables.

Note that in cases where the core is empty, we can relax the core constraint to $d^C \leq \varepsilon$, where $\varepsilon$ is a small number that allows for a feasible solution. This variable could be minimized in the objective function of \ref{eq:wd}. As a result we would get a least core solution. Alternative solution concepts when the core is empty are the nucleolus or the kernel, but they are computationally more demanding such that we do not consider them further in this paper.

A key challenge for the implementation of \ref{eq:wd} is the fact that the number of follower constraints is exponential in the cardinality of the bidders' set. In addition, we have a number of bilinear terms such as $p_i(S)x_i(S)$ and $p_j(Z)y_j(Z)$ which need to be linearized.
Note that the bi-linear term $\sum_{S \subseteq K} p_{i}(S) x_i(S)$ in constraints (BC) and (BB) can easily be replaced by a single variable $p_i$ ($p_j$) for non-linear and personalized prices. In contrast, sometimes an auctioneer might want to have non-linear but anonymous prices and he could replace the variables $p_{i}(S)$ ($p_{j}(Z)$) for all $i$ ($j$) by a single variable $p(S)$ ($p(Z)$) for each package $S$ ($Z$). Note that neither personalized nor anonymous prices might be unique.

\subsection{Calculating the least core}

Unfortunately, the core of the combinatorial exchange can also be empty, i.e. there may exist no allocation of items with prices for which there is no blocking coalition. However, the gains for each coalition might only be marginal and exceed the costs of finding such a blocking coalition for the participants. In cooperative game theory, the least core is defined as the set of outcomes for which the maximum profit of a blocking coalition is minimal. For the combinatorial combinatorial exchange as defined above, this corresponds to minimizing the maximal improvement of the coalitions in comparison to the upper level allocation and prices. Then, instead of maximizing the gains from trade, a least core assignment can be determined by defining the maximal improvement $\Delta \ge \max\left\{0, \max_C d^C \right\}$ and replacing the objective function by $\min \Delta$. Another possibility is to define a combination of the objective functions, assigning weights to the gains from trade and the maximal possible improvement.

\section{Algorithmic Approaches to the General Problem}

In this section, we discuss algorithmic approaches to solving MIBLPs in general. Then, we introduce computations allowing us to reduce the number of coalitions we need to consider in the general case and a decomposition approach that reduces the problem size significantly in practice. 

\subsection{Mixed Integer Bilevel Linear Programs}

Bilevel optimization has its roots in the seminal work by \cite{von1934marktform}. Bilevel linear programs (BLPs) are already $NP$-hard, and integer bilevel programs (IBLPs) are $\Sigma_2^p$-complete \citep{jeroslow1985polynomial}. For the following discussion, we introduce the short hand of a generic mixed integer bilevel program (MIBLP):

\begin{align*}
	\textrm{max} & F(x,y) \\
	\textrm{s.t.} & G(x,y) \le 0 \\
	& y \in \textrm{argmin}\left\{f(x,y'), \textrm{s.t.} g(x,y') \le 0, y' \in Y\right\} \\
	& x \in X
\end{align*}

Here, $F,f,G$ and $g$ are linear functions and $X,Y \subset \mathbb{Z} \times \mathbb{R}$ are the respective domains of upper-level variables $x$ and lower-level variables $y$.

%
%



\cite{bard1990branch} initiated algorithmic solutions to mixed integer bilevel linear programs (MIBLPs). Their algorithm converges if either all leader variables are integer, or when the follower subproblem is an LP. Until recently, MIBLPs were considered ''still unsolved by the operations research community'' \citep{delgadillo2010analysis}. Only this year, two general purpose branch-and-cut MIBLP algorithms have been proposed by \cite{fischetti2017new} and \cite{tahernejadbranch}. \cite{fischetti2017new} extend their earlier algorithm for MIBLPs with binary first-level variables to problems where linking variables are discrete. In a very recent unpublished paper, \cite{tahernejadbranch} propose another general-purpose MIBLP solver based on branch-and-cut which is available open source in the MibS solver. The latter requires the linking variables, those variables that have non-zero coefficients and are present in the upper- and lower-level program, to be integer. Since for our domain, the linking variables contain (possibly non-integer) upper level prices, we implemented the column-and-constraint generation algorithm proposed by \cite{zeng2014solving}, which is applicable to general mixed-integer bilevel problems. 

In the following, we give a short outline of the algorithm by \cite{zeng2014solving}. First, a single-level reformulation of the bilevel program is introduced, wherein all lower level variables and constraints are duplicated into the upper level and a classical MILP is solved which yields a solution that is feasible with respect to upper and lower level constraints. However, assignment of the lower level variables does not necessarily yield an optimal solution for the lower level problem and the solution of this relaxation only serves as an upper bound $UB$ for the MIBLP. Given an optimal assignment $x^*$ of the upper level variables in the single-level reformulation, the lower level problem is then solved to optimality, yielding an assignment $y^*$ for the lower level variables. If the combined solution $(x^*,y^*)$ is feasible for the MILBP, then $F(x^*,y^*)$ is a lower bound $LB$ for its optimal solution. In the case that $LB = UB$, $(x^*,y^*)$ is also an optimal solution. Otherwise, let $y_{\mathbb{Z}} \in Y_{\mathbb{Z}}$ consist of the lower level variables with integer domain and $y_{\mathbb{R}} \in Y_{\mathbb{R}}$ denote the continuous lower level variables. The single-level reformulation is extended by the Karush-Kuhn-Tucker (KKT) optimality conditions of the lower level with the integer variables fixed to $y^*_{\mathbb{Z}}$. The procedure continues as described above, until lower bound and upper bound converge to the same value or the single-level reformulation is infeasible. While in the worst case, the algorithm requires enumeration of all possible assignments of the lower level integer variables, in practice it converges in only a few iterations.




\subsection{Introducing Cuts to Reduce the Number of Coalitions}

Our goal is to find the core allocation that maximizes welfare. A few problem specifics raise hope that we can reduce the number of coalitions significantly. 
First, we introduce \eqref{eq:wdb} a program to compute allocations based on the valuations capped by the budget constraint.
	
	\begin{small}
\[\begin{array}{rcll@{\hspace{2em}}l} 
w_B=\max & \sum_{i\in I}\sum_{S \subseteq K} \min\left\{B_i,v_{i}(S)\right\}x_{i}(S) - \sum_{j\in J}\sum_{Z \subseteq K}v_{j}(Z)y_{j}(Z)& \label{eq:wdb} \tag{\textbf{B}}&  \\
\textrm{s.t.} & x_{i}(S)-\sum_{x: x_{i}=S}\delta_{X,Y}=0 & \forall i\in I,\forall S\subseteq K & (p_{i}(S)) \\
 & -y_{j}(Z)+\sum_{y: y_{j}=Z}\delta_{X,Y}=0 & \forall j\in J,\forall Z\subseteq K & (p_{j}(Z)) \\
 & \sum_{S\subseteq K}x_{i}(S)\leq 1 & \forall i \in I & (\pi_{i})\\
 & \sum_{Z\subseteq K}y_{j}(Z)\leq 1 & \forall j \in J & (\pi_{j})\\
 & \sum_{(X,Y)\in\Gamma}\delta_{X,Y} = 1 & & ( \pi_a)\\
 & 0\leq x_{i}(S) & \forall S\subseteq K,\forall i\in I&\\
 & 0\leq y_{j}(Z) & \forall S\subseteq K,\forall j\in J&\\
 & 0\leq\delta_{X,Y} & \forall (X,Y)\in\Gamma&\\
 \end{array}\]
\end{small}
	
The allocation from \eqref{eq:wdb} does not need to be the one that maximizes welfare, but model \eqref{eq:wdb} can be instrumental to cut the number of coalitions in \eqref{eq:wd}. 

\hide{
\begin{example}
Suppose we have four bidders with the valuations and budget constraints for items $A$, $B$, and the package $\{A,B\}$ described in Table \ref{welfaretheo1}. Giving item $A$ to $b_1$ and $B$ to $b_2$ for a price of \$2 each is optimal for \eqref{eq:wdb}, but there is another core allocation where the package $\{A,B\}$ is assigned to bidder $b_1$ for a price of \$4, or $A$ could be assigned to bidder $b_4$ and $B$ to bidder $b_3$ with the same objective function value in \eqref{eq:wdb}. 

\begin{table}[ht]
\centering
\caption{Example with budget constraints $B_i$}
\label{welfaretheo1}
\begin{tabular}{lllll}
\hline
     & \{A\} & \{B\} & \{A,B\} & $B_i$\\ \hline
 $b_1$  & 2 &   & 10 & 4 \\
 $b_2$  &    & 2 & 2 & 2 \\
 $b_3$  &    & 4 & 4 & 2 \\
 $b_4$  & 6  &   & 6 & 2 \\ \hline
\end{tabular}
\end{table}
\end{example}


} %

We use $\alpha$ as a short hand to describe the optimal allocation $\left(X(\alpha),Y(\alpha)\right)$ resulting from \eqref{eq:wdp} and $\beta$ to denote the optimal allocation resulting from \eqref{eq:wdb}. We use $C_{\alpha}$ to describe the corresponding coalition of bidders in allocation $\alpha$.

\begin{theorem}\label{prop:capped}
If \eqref{eq:wdp} is feasible, then the capped coalitional value $w_B(\alpha)$ of an optimal allocation $\alpha$ computed by \eqref{eq:wdp} cannot be higher than the capped coalitional value $w_B(\beta)$ of an optimal allocation $\beta$ for \eqref{eq:wdb}, i.e  $w_B(\alpha) \leq w_B(\beta)$. 
\end{theorem}

\proof{Proof:}
Allocation $\alpha = \left(X(\alpha),Y(\alpha)\right)$ is a feasible for \eqref{eq:wdp}. 
Hence, $\left(X(\alpha),Y(\alpha)\right) \in \Gamma$ in \eqref{eq:wdb} and $X(\alpha), Y(\alpha), \delta_{X(\alpha),Y(\alpha)}$ is feasible for \eqref{eq:wdb}. Then, the result immediately follows, since $\beta$ is optimal for \eqref{eq:wdb}. \qed 
\endproof

Note that it can well happen that for an optimal allocation $\alpha$ of \eqref{eq:wdp} and an optimal allocation $\beta$ of \eqref{eq:wdb} it holds that $w_B(\alpha) < w_B(\beta)$, although $w_P(\alpha) > w_P(\beta)$.

\begin{example}
Suppose we have a market with two buyers and two sellers selling one of two items each as described in Table \ref{ex:core}. The sellers have zero value for the objects in this example. An optimal solution $\beta$ of \eqref{eq:wdb} would be to assign $A$ to $b_1$ for a price of zero and $B$ to $b_2$ for a price of \$2 with $w_B(\beta) = w_P(\beta) = 4$. This allocation would not be in the core, because bidder $b_1$ would prefer item $B$ and be willing to pay a price of say \$2.5 to the seller, the coalition of $b_1$ and the seller of $B$ is a blocking coalition. In \eqref{eq:wdp}, an optimal solution $\alpha$ would consist of $b_1$ receiving $B$ for a price of $2$, for example, resulting in $w_P(\alpha) = 10 > w_P(\beta) = 4$ but at the same time $w_B(\alpha) = 3 < w_B(\beta)=4$. 

\begin{table}[ht]
\centering
\caption{Example}
\label{ex:core}
\begin{tabular}{llll}
\hline
 & \{A\} & \{B\} & $B_i$ \\
 $b_1$  & 2  & 10 & 3 \\
 $b_2$  &    & 2 & 2  \\ \hline
\end{tabular}
\end{table}
\end{example}

Still, this observation can vastly reduce the number of coalitions one needs to explore, because we can ignore coalitions $C_{\alpha}$ where $w_P(\alpha) < w_P(\beta)$ in \eqref{eq:wd}. This means, before setting up \eqref{eq:wd}, one can compute \eqref{eq:wdb} and \eqref{eq:wdp} for all coalitions and omit those where $w_P(\alpha) < w_P(\beta)$ in the \eqref{eq:wd}.


Note that \eqref{eq:wdb} has a variable $\delta_{X,Y}$ for each allocation. While this formulation is a linear program, the number of variables might not be practical. Instead of \eqref{eq:wdb}, we can also compute the upper-level program as a set packing problem, capping $v_i(S)$ by $\min\{B_i,v_{i}(S)\}$ to get $w_B(\beta)$.  

\subsection{Decomposition by Delayed Coalition Generation}

Another approach to cope with the exponential number of coalitions in the MIBLP is the delayed generation of coalitions. Instead of determining an allocation and corresponding prices which are in the core with respect to all possible coalitions, we determine an initial set of coalitions $\mathcal{C} \subseteq C$ and solve the MIBLP, only considering coalitions in $\mathcal{C}$. 

For example, this initial set $\mathcal{C}$ can be a set of smaller coalitions with high gains from trade that are sufficiently different from the grand coalition, because they might be able to redistribute their gains from trade in a way that is better for all of them. These coalitions have a higher likelihood of becoming a blocking coalition. The auctioneer can compute the optimal allocation of all smaller coalitions $\hat{\mathcal{C}_l} \subseteq C$ up to a limited number of participants $l$. Computing the allocation problem (P) for small coalitions of only a few bidders $c \in \hat{\mathcal{C}_l}$ can be done fast in practice. Then, the auctioneer selects those coalitions with a high objective function value and a high Hamming distance between the optimal allocation of (P) considering the grand coalition $C$ and that of the a small coalition $c \in \hat{\mathcal{C}_l}$ and add it to $\mathcal{C}$.

The gains from trade we obtain for the allocation $(x^*,y*)$ and prices $p^*$ that are feasible with regards to the set $\mathcal{C}$ serve as an upper bound for the complete problem considering all possible coalitions. Moreover, in the case that there is no core outcome for the set $\mathcal{C}$, there is also no core outcome for the complete problem. 

To check whether the allocation $(x^*,y^*)$ and prices $p^*$ are feasible with regards to all other coalitions $C \setminus \mathcal{C}$, we only need to solve the lower level problem (i.e. an integer program) for these coalitions. Suppose, there is a coalition $C_0$ with $d^{C_0} > 0$, then the solution of the MIBLP for coalitions $\mathcal{C}$ is not in the core for all coalitions $C$. In this case, we extend the set $\mathcal{C}$ to the set $\mathcal{C}' = \mathcal{C} \cup \left\{ C_0 \right\}$ and solve the MIBLP again, considering coalitions $\mathcal{C}'$. On the other hand, if there is no coalition $C_0 \in C \setminus \mathcal{C}$ with $d^{C_0} > 0$, the solution of the MIBLP for coalitions $\mathcal{C}$ is in the core when considering all coalitions and thus yields an allocation and prices with maximal gains from trade. Then, we are done. 
Using this decomposition approach employed in our numerical experiments, we could often significantly reduce the size of the MIBLPs that needed to be solved.

\section{Restricted Cases}
\label{sec:restricted}

The complexity analysis of the allocation problem in combinatorial auctions has drawn considerable attention and led to a characterization of tractable cases where the LP relaxation is integral \citep{muller2006tractable}. These cases often depend on the types of valuations which are typically unknown ex ante. Total unimodularity of the constraint matrix of the allocation problem or substitutes valuations are an example for tractable cases. 

In this section we analyze important special cases of the problem \eqref{eq:wd}, which are simple to characterize ex ante and not $\Sigma_2^p$-hard. A simple case is obviously when none of the budget constraints is binding, which leads to the traditional case with quasi-linear utilities \citep{bichler2017core}. Another extreme case is where the budget constraints are all zero such that with a given set of sellers who have some positive value for the good, there would be no trade. 


In the following, we discuss single-sided combinatorial auctions with non-zero, but binding budget constraints, and combinatorial exchanges where we only care about blocking dyadic coalitions. Both allow for computations that are $NP$-hard, but not $\Sigma_2^p$-hard.

\subsection{Restricting to One-Sided Auctions}

If we had a combinatorial auction with only a single seller (or only a single buyer, resp.), we are able to decouple the allocation and pricing problem. With a single seller, we can solve \eqref{eq:wdb}. This allocation maximizes the revenue of the seller such that he does not have an incentive to deviate. In case of a unique optimal allocation, buyers with a high value but low budget cannot deviate and make themselves and the seller better of. The concept of the (weak) core requires that there does not exist a coalition of buyers and the seller such that they all (strictly) prefer an alternative allocation. 

The integer program \eqref{eq:wdca} replaces the LP \eqref{eq:wdb} with the large number of variables $\delta_{X,Y}$ and computes an allocation that maximizes seller revenue. Subsequently, we can compute core prices based on the capped valuations following the algorithms suggested by \cite{Day07b} or \cite{erdil2010new}. In such an allocation, the seller cannot improve his utility (i.e. revenue) strictly yielding a weak core solution.

\begin{subequations}
\begin{align} 
   z^* = \max_{x_{i}(S)} & \sum_{S \subseteq K} \sum_{i \in I} \min{(B_i,v_{i}(S))} x_i(S) & \label{eq:wdca}  \tag{\textbf{CA-S}} \\
  \text{subject to }  \notag \\
	&\sum_{S: k \in K} \sum_{i \in I} x_i(S) \leq 1 & \forall k \in K \label{Supply} \tag{Supply}\\
  &\sum_{S \subseteq K} x_i(S) \leq 1 & \forall i \in I \label{xor} \tag{XOR}\\
	& x_{i}(S) \in \{0,1\} & \forall S \subseteq K, i \in I \label{wd6} \tag{Binary}
\end{align}
\end{subequations}

After computing \eqref{eq:wdca}, there is still a possibility that there is another stable allocation with the same objective function value or seller revenue, but a higher payoff for the buyers based on their uncapped valuations. This means that we can achieve a weak core outcome with a higher welfare, which is also our goal in \eqref{eq:wd}. For this, we solve a second optimization problem  \eqref{eq:wdcab} before the core-price computations, in order to get weak core outcomes. 

\begin{subequations}
\begin{align} 
   \max_{x_{i}(S)} & \sum_{S \subseteq K} \sum_{i \in I} v_{i}(S)x_i(S) &  \label{eq:wdcab} \tag{\textbf{CA-B}} \\
  \text{subject to }  \notag \\
	& \sum_{S \subseteq K} \sum_{i \in I} \min{(B_i,v_{i}(S))} x_i(S) \geq z^* \notag \\
	&\sum_{S: k \in K} \sum_{i \in I} x_i(S) \leq 1 & \forall k \in K \label{Supply} \tag{Supply}\\
  &\sum_{S \subseteq K} x_i(S) \leq 1 & \forall i \in I \label{xor} \tag{XOR}\\
	& x_{i}(S) \in \{0,1\} & \forall S \subseteq K, i \in I \label{wd6} \tag{Binary}
\end{align}
\end{subequations}

Note that the \textit{strong core} refers to an outcome, where there is no coalition that could make all its members at least as good and at least one member better off. In single-sided auctions the strong core can be empty. To see this, consider an auction with a single object and two bidders with the same value $v$. In the first allocation, the seller sells the object to buyer 1 at price $p \leq v$, in the second allocation he sells to bidder 2 at price $p$. In both cases, the revenue of the auctioneer does not increase, but there is one bidder, whose payoff would increase strictly. 
Similarly, suppose that we get a winning coalition $C \subset I$ from \eqref{eq:wdca} and another winning coalition $C'$ with a different set of buyers after we compute \eqref{eq:wdcab}. Now, if we switch back from $C'$ to coalition $C$ this set of buyers improves payoff while the revenue of the seller remains the same.

\subsection{Restricting the Size of Coalitions}

The concept of the core considers coalitions of any size. Large coalitions are costly to form. However, it is simpler and therefore more likely to find blocking pairs of one buyer and one seller only. For some applications, it might be sufficient to find a solution that avoids deviations of dyadic coalitions. We do so with the mixed binary program \eqref{eq:wddy}. 


For this, we introduce variables $\rho_{ij}(S) \in \{0,1\}$ for each possible package trade between a buyer $i$ and seller $j$ over all packages $S \subset K(j)$ where $K(j)$ denotes all bundles $Z \subseteq K$ which are offered by $j$. The variable $\rho$ is set to $1$, whenever a dyadic coalition would form a blocking coalition. Similar to the general problem, we introduce constraints such that only outcomes without blocking coalitions are feasible.

Constraints \eqref{eq:c1} to \eqref{eq:c3} characterize blocking dyads and require some explanation. Note that a buyer $i$ would want to deviate if his payoff $v_i(S)-p_{ij}(S)>\pi_i$ where $p_{ij}(S)$ is some transfer price in a blocking dyad.  Similarly, a seller $j$ would want to deviate if $p_{ij}(S) - v_j(S) > \pi_j$. Rearranging terms, $\pi_j + v_j(S) < v_i(S) - \pi_i$ characterizes a blocking coalition, i.e. with $\pi_j + v_j(S) \geq v_i(S) - \pi_i$ a dyad would not be blocking (see \eqref{eq:c2}). We also need to consider budget constraints of buyers $B_i$. With $\pi_j + v_j(S) > B_i$ in constraint \eqref{eq:c1}, we avoid payments to the seller $j$ characterized by the LHS of the constraint that are higher than the budget of the buyer $B_i$. The binary variable $\gamma_{ij}(S)=1$ indicates if a dyad would deviate due to improvement in payoffs, variable $\delta_{ij}(S)=1$ if the required payments would exceed budget. Constraint \eqref{eq:c3} demands that a dyad can only be willing to deviate due to payoffs if the required payments would exceed the budget of the buyer involved, since otherwise this dyad would be blocking the outcome.


\begin{tiny}
\begin{subequations}
\begin{align} 
 \max_{x_{i}(S), y_j(Z)} & \sum_{S \subseteq K} \sum_{i \in I} v_{i}(S) x_i(S)  - \sum_{j\in J}\sum_{Z \subseteq K}v_{j}(Z)y_{j}(Z) \label{eq:wddy} \rho_{ijS} \tag{\textbf{DY}} \\
  \text{s.t.}  &\sum_{S \subseteq K} p_{i}(S) x_i(S) \leq \min\left\{B_i,\sum_{S \subseteq K} v_{i}(S) x_i(S)\right\}  & \forall i \in I \label{eq:ul_bc} \tag{BC}\\\
	&p_{j}(Z) y_j(Z) \geq v_{j}(Z) y_j(Z) \label{eq:ul_bb} & \forall Z \subseteq K, \forall j \in J \tag{IRS}\\	
	&\sum_{i \in I} \sum_{S \subseteq K} p_{i}(S) x_i(S) = \sum_{j \in J} \sum_{Z \subseteq K} p_{j}(Z) y_j(Z) \label{eq:ul_bb} \tag{BB}\\
	&\sum_{S: k \in K} \sum_{i \in I} x_i(S) \leq \sum_{Z: k \in K} \sum_{j \in J} y_j(Z) & \forall k \in K \label{eq:ul_supply} \tag{Supply}\\
	&\sum_{S \subseteq K} x_i(S) \leq 1 & \forall i \in I \label{eq:ul_xor-b} \tag{XOR-B}\\
	&\sum_{Z \subseteq K} y_j(Z) \leq 1 & \forall j \in J \label{eq:ul_xor-s} \tag{XOR-S}\\
	&\pi_i = \sum_{S \subseteq K} (v_i(S)-p_i(S))x_i(S) & \forall i \in I \label{eq:payoff} \tag{payoffB}\\
	&\pi_j = \sum_{Z \subseteq K} (p_j(Z)-v_j(Z))y_j(Z) & \forall j \in J  \label{eq:payoff} \tag{payoffS}\\
  &\pi_{j} + v_{j}(S) \ge B_i\delta_{ij}(S) & \forall i \in I, \forall j \in J, \forall S \subseteq K(j)  \label{eq:c1} \tag{Block-B} \\
  &\pi_{j} + v_{j}(S) \ge v_i(S) - \pi_i - M\gamma_{ij}(S) & \forall i \in I, \forall j \in J, \forall S \subseteq K(j)  \label{eq:c2} \tag{Block-Imp} \\
  &\delta_{ij}(S) \ge \gamma_{ij}(S) & \forall i \in I, \forall j \in J, \forall S \subseteq K(j)   \label{eq:c3} \tag{No-Block}\\
	& x_{i}(S) \in \{0,1\} & \forall S \subseteq K, i \in I \label{leader_b} \tag{Binary}\\
	& y_{j}(Z) \in \{0,1\} & \forall Z \subseteq K, j \in J \label{leader_b2} \tag{Binary}\\
  & \delta_{ij}(S), \gamma_{ij}(S), \rho_{ij}(S) \in \{0,1\} & \forall i \in I,  \forall j \in J, \forall S \subseteq K(j) \tag{Binary}\\
	& \pi_i, p_{i}(S) \in \mathbb{R}^+_0 & \forall S \subseteq K, i \in I  \label{leader_r} \tag{Real}\\
	& \pi_j, p_{j}(Z) \in \mathbb{R}^+_0 & \forall Z \subseteq K, j \in J  \label{leader_r2} \tag{Real}
\end{align}
\end{subequations}
\end{tiny}

Beyond dyadic coalitions, one could restrict the cardinality of coalitions to those with only three or four participants. Even the computation of the coalitional value of these small coalitions is NP-hard in general. However, such a restriction on the coalitions reduces the number of lower-level programs in \eqref{eq:wd} from $2^{|K|}-1$ to $\sum_{i=1}^k$ ${|K|}\choose{i}$ with $k$ being the maximum size of the coalition.

\section{Experimental Results}
\label{sec:results}

Even though the problem of finding a core allocation is computationally very hard to solve exactly in the worst case, it can still be possible to solve problem sizes that are practically relevant. We provide experimental results suggesting that the computation of core outcomes in the presence of budget constraints might well be possible for restricted problem sizes even on commodity hardware. We ran a number of experiments on a standard laptop with an Intel Core 17-7600U CPU (2.9 GHz) with 16 GB memory on a 64-bit Windows operating system. Our implementation for the MIBLP is based on that of \cite{zeng2014solving} extended by the decomposition by delayed coalition generation described in the previous section.

\subsection{Data}

In our introduction, we have described combinatorial exchanges for fishery access rights and airport time slots as motivating examples. Unfortunately, there are no publicly available datasets of combinatorial exchanges that we are aware of. However, the Combinatorial Auctions Test Suite (CATS) \citep{leyton2000towards} is the most widely used benchmark for the evaluation of algorithms for the combinatorial auction problem. One of the CATS instance generators models the airport time slot problem, which can be seen as a combinatorial exchange mechanism among airlines. This allows us to provide experimental results that can be replicated by others.

This instance generator models the four largest USA airports, each having a predefined number of departure and arrival time slots. 
For simplicity there is only one slot for each time unit available. Each bidder is interested in obtaining one departure and one arrival slot (i.e., item) in two randomly selected airports. His valuation is proportional to the distance between the airports and reaches a maximum when the arrival time matches a certain randomly selected value. The valuation is reduced if the arrival time deviates from this optimal value, or if the time between departure and arrival slots is longer than necessary. Further, we extended the CATS generator to include budget constraints for bidders. For each bidder, her budget is generated by a random uniform draw from the interval of zero to her maximal valuation for any of her desired item.

\subsection{Results}

We report the results of the allocation and pricing problem in the combinatorial exchange of airport time slots created with the CATS instance generator. Treatment variables include the number of airlines (referred to as \textit{bidders}) and the number of time slots (referred to as \textit{items}) which we distributed evenly across the four \textit{airports}. Further, we evaluated the outcomes when we only consider blocking coalitions of three or five participants instead of blocking coalitions of unrestricted size. We will refer to such outcomes where we only check for blocking coalitions with at most $n$ bidders as $n$-core. Given that it is computationally very hard to find deviating coalitions, an outcome that is stable against coalitions of smaller size might be a sufficient stability notion in practice.

Table \ref{tab:results} shows the average results of 50 random instances for each of the treatment combinations. We report the number of items and bidders as parameters for the problem size, as well as the maximum size of the coalitions we consider. The number of coalitions up to a specific size is shown in Table \ref{tab:numcoals}. For each treatment combination we report how many of the 50 instances could be solved within five minutes (i.e. either a welfare-maximizing core outcome was found or the core was proven to be empty) as well as the average runtime of those instances for which our MIBLP implementation terminates within 5 minutes. 

\begin{table}[h]
\centering
\begin{tabular}{l|r|r|r}
$|I|$ & size 3 & size 5 & unrestricted size \\
3 & 18 & 54 & 59\\ 
7 & 42 & 462 & 1,341  \\ 
11 & 66 & 1,606 & 22,407\\ 
\end{tabular}
\caption{Number of coalitions of up to a specific size for the problem sets.}
\label{tab:numcoals}
\end{table}

\begin{table}[h]
\centering
\begin{tabular}{ll|lr|lr|lr}
 & & \multicolumn{2}{c}{3-core} &  \multicolumn{2}{c}{5-core} &  \multicolumn{2}{c}{unb. coal. size} \\
$|I|$ & $|\mathcal{K}|$ & solved & avg. runtime & solved & avg. runtime & solved & avg. runtime \\ \hline
3 & 6  & 50 & 0.10 & 50 & 0.52	& 50	& 0.55  \\ 
3 & 12 & 50 & 0.05 & 50 & 0.40	& 50	& 0.52  \\ 
3 & 18 & 50 & 0.06 & 50 & 0.22	& 50	& 0.35  \\ 
3 & 24 & 50 & 0.10 & 50 & 0.55	& 50	& 1.03  \\ 
7 & 6  & 50 & 0.30 & 50 & 0.75	& 50	& 1.82  \\ 
7 & 12 & 50 & 0.20 & 50 & 8.96	& 36	& 206.64\\ 
7 & 18 & 50 & 0.36 & 48 & 9.28	& 30	& 385.48\\ 
7 & 24 & 50 & 1.27 & 50 & 26.78	& 33	& 404.63\\ 
11& 6  & 50 & 1.03 & 50 & 6.20	& 50	& 49.04 \\ 
11& 12 & 50 & 1.29 & 41 & 48.90 & 8		& 211.98\\ 
11& 18 & 50 & 2.51 & 47 & 48.91	& 5		& 238.24\\ 
11& 24 & 50 & 3.65 & 48 & 53.64	& 3		& 181.63\\ 
\end{tabular}
\caption{Solved instances and average runtimes in seconds of 50 random instances}
\label{tab:results}
\end{table}

It can be seen that the problem sets with three or seven bidders can all be solved within an average of one to two seconds even when considering all possible coalitions. Even for 11 bidders all instances of the problem set with 6 items can be solved within an average of 50 seconds. For 7 bidders, we were still able to solve at least 30 of the 50 instances for each treatment combination. Even for the largest problem sets, it was possible to obtain solutions within five minutes on commodity hardware. Note that when considering only coalitions of size up to 3 or 5, almost all instances could be solved. We have taken the largest instances with 11 bidders and 24 items and run them with a time limit of 30 minutes. While the results did not change significantly, this allowed us to find a 5-core solution for the two remaining instances and obtain solutions for five additional instances without restrictions of the core. 

In Table \ref{tab:feas}, we report the number of instances where the core is not empty when considering three, five, or an unbounded size of blocking coalitions. Further, we show how many of the welfare-optimal core outcomes that were computed on the basis of being stable against blocking coalitions of size at most three (3-core) are not even blocked by coalitions of larger sizes (in 5-core or in core). Similarly, we also report the number of coalitions that are in the 5-core and how many of them are in the core, i.e. stable against all possible coalitions. Note that this is straightforward to check by  evaluating the allocation and prices in the 5-core against all possible coalitions, for example. Such outcomes are also welfare-optimal core outcomes when considering these coalitions. 

Interestingly, in the largest problem set with 11 bidders and 24 items, 8 of the outcomes that are in the 5-core are in the core in general. This is more than the number of core outcomes that we could compute in time in via the MIBLP approach (see Table \ref{tab:results}). In other words, an $n$-core solution can often be a core solution even if the core outcome is intractable.

\begin{table}[h]
\centering
\begin{tabular}{ll|lll|ll|l}
 & & \multicolumn{3}{c}{3-core} &  \multicolumn{2}{c}{5-core} &  {core} \\
$|I|$ & $|\mathcal{K}|$ & in 3-core & in 5-core & in core & in 5-core & in core & in core\\ \hline
3 & 6  & 47	& 47	& 47	& 47	& 47	& 47	\\ 
3 & 12 & 50	& 35	& 35	& 50	& 50	& 50	\\ 
3 & 18 & 50	& 42	& 42	& 50	& 50	& 50	\\ 
3 & 24 & 50	& 45	& 45	& 50	& 50	& 50	\\ 
7 & 6  & 48	& 48	& 48	& 48	& 48	& 48	\\ 
7 & 12 & 50	& 6		& 3		& 50	& 23	& 36	\\ 
7 & 18 & 50	& 11	& 5		& 48	& 13	& 30	\\ 
7 & 24 & 50	& 16	& 8		& 50	& 21	& 33	\\ 
11& 6  & 47	& 47	& 47	& 47	& 47	& 47	\\ 
11& 12 & 50	& 2		& 0		& 41	& 10	& 8		\\ 
11& 18 & 50	& 8		& 2		& 47	& 4		& 5		\\ 
11& 24 & 50	& 8		& 2		& 48	& 8		& 3		\\ 
\end{tabular}
\caption{Number of instances with a welfare-maximal outcome that are in 3-core, 5-core, and in the core wrt. all possible coalitions. Additionally, number of these outcomes that are in the core for coalitions of larger sizes.}
\label{tab:feas}
\end{table}

A more extensive experimental evaluation with different types of problem instances is beyond the scope of this paper. However, the results indicate that even though the general problem of finding a core-stable outcome in a combinatorial exchange with budget constraints is $\Sigma_2^p$-complete, computing $n$-core outcomes can be a viable approach in practice. 
\section{Conclusions}

We analyze combinatorial exchanges in the presence of financially constrained bidders. 
Our analysis shows that budget constraints lead to additional core constraints in the allocation problem, and that as a result computing allocation and prices becomes $\Sigma_2^p$-hard. We introduce mathematical optimization problems effective computational techniques to solve these problems for restricted problem sizes in practice. Even if we could only compute core-stable outcomes that are stable against small but not all  coalitions, this might provide a sufficient level of stability in practice. This is, because it will be as difficult for participants as it is for the auctioneer to find a blocking coalition. 

We emphasized stability over other design desiderata in our paper, and this deserves some discussion. 
Even in those cases where we can compute a core-stable allocation considering the budget constraints and valuations of bidders, it is clear that these financial constraints restrict the possible gains from trade and the allocation will likely be different and with lower welfare than an allocation not considering budget constraints and just the valuations. 
However, in a matching market without money if sellers had some positive value for their goods and buyers are not able to pay, then no trade would take place and no gains from trade would emerge. 




Note that stability is a first-order design goal also in other markets, even if it is at the expense of allocative efficiency. For example, it is well-known that the deferred acceptance algorithm by \cite{gale1962college} is stable but not efficient, while the top trading cycles algorithm by \cite{shapley1974cores} is efficient, but not stable. 
Later, \cite{roth2002economist} showed on the basis of empirical observations that stability is a key feature of successful matching mechanisms in practice. 
We argue that the stability of a market outcome is also important in price-guided markets such as that for fishery access rights discussed in the introduction. 

\bibliographystyle{ormsv080} 
\bibliography{BibFile} 


\newpage
\begin{APPENDICES}
\section{Complexity Analysis}
\label{sec:complex}

In the following we prove that finding a welfare-maximizing core allocation with exogenous budget constraints is $\Sigma_2^p$-complete by a reduction from the canonical $\Sigma_2^p$-complete problem QSAT$_2$.

\textbf{2-Quantified Satisfiability, QSAT$_2$}:
Given a $n+m$ variable Boolean formula $\varphi(x,y)$ in DNF with $x=\left(x_1,\ldots,x_n\right)$ and $y = \left(y_1,\ldots,y_m\right)$ is it true that $\exists x \forall y \varphi(x,y)$?

\subsection{Membership in $\Sigma_2^p$}

We first prove that the problem of finding a core outcome of welfare $D$ is in the class $\Sigma_2^p$. Let $x(S),y(Z),p(S),p(Z)$ be a certificate for the allocations and prices. The gains from trades can be easily verified in polynomial time by using this certificate. Further, showing that this outcome is in the core is in $co-NP$ since any blocking coalition $C$ with corresponding assignments $\chi^C(S),\gamma^C(Z),p^C(S),p^C(Z)$ is a certificate that the outcome is not in the core.

\subsection{Idea behind the transformation}

Before formally proving the theorem, we give a short explanation of the reduction and the ensuing relationship between an instance of QSAT$_2$ and the corresponding combinatorial exchange. We concentrate on the main items and buyers with a direct correspondence to the world of QSAT$_2$ and omit the various auxiliary items, buyers and sellers. For these, we refer to the complete description of the transformation below.

In the combinatorial exchange, we define items relating to the truth assignment of $x$ and $y$ variables as well as the truth values which clauses evaluate to. For the variables $x$ and $y$, items $\chi$ and $\gamma$ are introduced and the truth assignment of variables $x$ and $y$ in QSAT$_2$ depends on which of the buyers obtains these items. Each clause is represented by $n^2$ items of type $\psi$ which will indicate whether the clause evaluates to true or false, again depending on which buyers obtain which of these items. 

We introduce different types of buyers. For $i \leq n$, buyers $B_i^{\mathcal{K}}$ and $B_i^{\mathcal{M}}$, each concerned with the items corresponding to the truth assignments of variable $x_i$ and the clauses affected by it. For $j \leq m$, buyers $B_j^{\mathcal{G}}$, which are concerned with the items corresponding to the truth assignments of $y_j$ and the clauses affected by it. The construction is such that either all buyers of type $B^{\mathcal{K}}$ win one of their preferred packages or they do not win any items. In the former case, the corresponding instance of QSAT$_2$ evaluates to true, in the latter case it is false. 

\begin{figure}
\begin{center}
\caption{Illustration of buyers' interests, only concerning items of type $\psi$}
\label{fig:trafo_idea}
\includegraphics[width=\columnwidth]{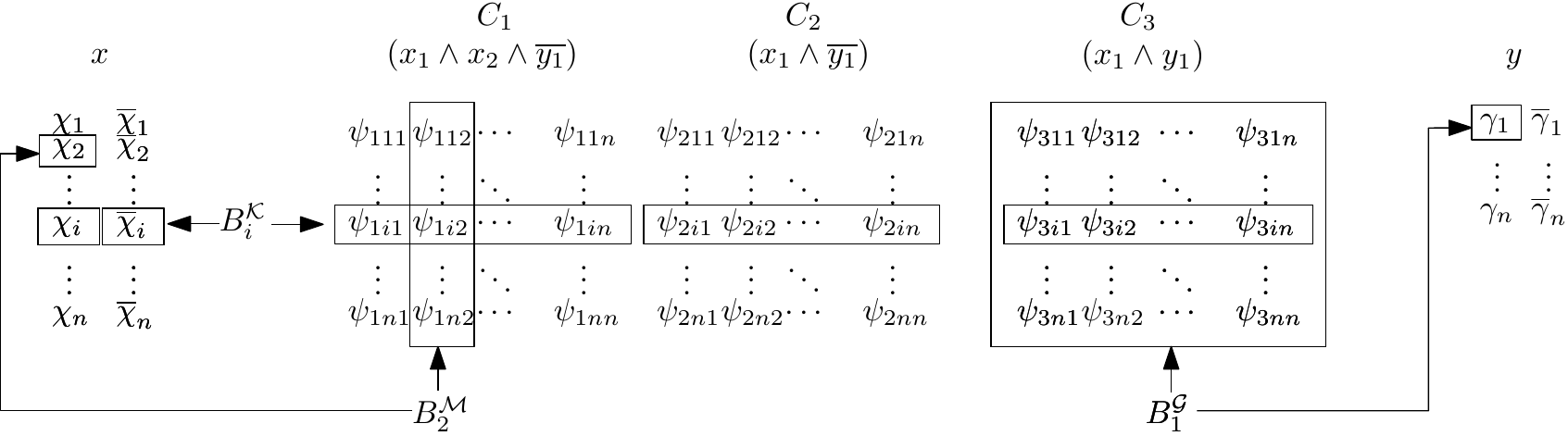}
\end{center}
\end{figure}

The connection between $\chi,\gamma$ and $\psi$ variables in the exchange and the correspondence of setting clauses to false by assigning truth values to variables in QSAT$_2$ is done via defining bundles of items in which the buyers are interested in. Figure \ref{fig:trafo_idea} demonstrates the situation, showing the items $\psi$ corresponding to three clauses in form of a matrix (we will refer to these as clause matrices in the following). Additionally, items $\chi$, $\gamma$ and bundles in which buyers of the various types are interested in, are shown. Buyer $B_i^{\mathcal{K}}$ is interested in either $\chi_i$ or $\overline{\chi}_i$ items as well as the $i$-th 'row' of one clause matrix. More formally, he is interested in the bundle 
\[
\left(\chi_i \vee \overline{\chi}_i\right) \wedge \left(\left\{\psi_{1i1},\ldots,\psi_{1in}\right\} \vee \left\{\psi_{2i1},\ldots,\psi_{2in}\right\} \vee \cdots \vee \left\{\psi_{Li1},\ldots,\psi_{Lin}\right\}\right),
\]
where $L$ is the number of clauses. Buyers $B_i^{\mathcal{M}}$ are interested in buying one out of $\chi_i$ or $\overline{\chi}_i$ as well as the $i$-th 'column' in all clause matrices of clauses which include the corresponding $x_i$ or $\overline{x_i}$ variable. In Figure \ref{fig:trafo_idea}, a bundle for buyer $B_2^{\mathcal{M}}$, including $\chi_2$ and the second column of the first clause matrix (since $C_1$ is the only clause containing $x_2$) is depicted. Finally, buyers $B_j^{\mathcal{G}}$ are interested in bundles which contain one item out of $\gamma_j$ or $\overline{\gamma}_j$ and complete clause matrices for clauses which include the corresponding $y_j$ or $\overline{y_j}$ variables. As can be seen, the individual bundles block each other and can not be obtained simultaneously for each clause matrix. The corresponding clause evaluates to true if and only if neither buyers of type $B^{\mathcal{M}}$ buys a column of the matrix or buyers of type $B^{\mathcal{G}}$ buy the complete matrix. For example, in Figure \ref{fig:trafo_idea}, no items of the second clause matrix are won by either a buyer type $B^\mathcal{M}$ or $B^\mathcal{G}$. In this case, buyers of type $B^{\mathcal{K}}$ can all obtain their respective row of the (second) clause matrix. Consequently, for this example, the second clause and therefore the entire expression evaluates to true.

The valuations and budgets of buyers are defined in such a way that buyers of type $B^{\mathcal{K}}$ have the highest value for their respective bundles, but only small budgets which does not allow them to bid up to their true valuation.  In contrast, buyers $B^{\mathcal{M}}$ have high valuations and sufficient budget to buy the bundle they are interested in. Buyers $B^{\mathcal{G}}$ have low valuations and can not compete with buyers $B^{\mathcal{M}}$. However, their budget is high enough in order to outbid buyers $B^{\mathcal{K}}$.
In order to obtain sufficiently high welfare gains, buyers $B^{\mathcal{K}}$ must obtain their desired bundles (i.e. win one of the clause matrices) and the outcome must be stable such that $B^{\mathcal{G}}$ and the sellers do not want to deviate by assigning the items to buyers $B^{\mathcal{G}}$ or $B^{\mathcal{M}}$ instead. Each buyer of type $B^{\mathcal{K}}$ can only obtain one of his desired bundles containing at least one row in one clause matrix (see Figure \ref{fig:trafo_idea}, which is equivalent to the corresponding clause evaluating to true in QSAT$_2$), when no other buyer purchases a column within this matrix. 

Buyers $B^{\mathcal{K}}$ and $B^{\mathcal{M}}$ are designed in such a way that $B_i^{\mathcal{K}}$ obtains the $\chi_i$-item corresponding to the truth assignment of $x_i$ and $B_i^{\mathcal{M}}$ its negation. Thus, buyers $B^{\mathcal{M}}$ obtain the columns in each clause matrix relating to the clauses which are set to false due to the truth assignment of variables $x$. Because of their lower valuations and budgets, buyers $B^{\mathcal{G}}$ can only compete for columns in clause matrices corresponding to clauses not yet set to false due to the assignment of $x$. These buyers maximize their payoffs when they can purchase as many complete matrices as possible which are not blocked by buyers $B^{\mathcal{M}}$. In QSAT$_2$ this corresponds to assigning truth values to variables $y$ in such a way that as many as possible of the remaining clauses evaluate to false (i.e. those which are not already evaluating to false due to the assignment of $x$ variables). Only if the buyers of type $B^{\mathcal{G}}$ cannot manage to block all remaining clause matrices (the $y$ variables in QSAT$_2$), buyers $B^{\mathcal{K}}$ can purchase rows in at least one of the matrices (the $x$ variables in QSAT$_2$) relating to one clause which evaluates to true. Then, the assignments of items corresponding to truth values of $x$ is a solution for the QSAT$_2$ problem. In other words, if  $B^{\mathcal{K}}$ win in every allocation, then there exists a stable outcome that achieves the pre-defined welfare in the decision problem.

\subsection{Transformation}

We present a transformation with valuations using an XOR bidding language. The transformation can easily be done for an OR bidding language as well, however this requires additional auxiliary items.

For a given formula $\varphi(x,y)$ with clauses $C_1,\ldots,C_L$ construct an instance $\textrm{CEx}_{\varphi(x,y)}$ of a combinatorial exchange with bidders and items as follows. First, consider $n + L + 2$ sellers:
\begin{itemize}
	\item One seller $S^\chi_i$ for each $i = 1,\ldots,n$. Each seller $S^\chi_i$ offers items $\chi_i$ and $\overline{\chi}_i$. These items will later indicate which logical values have to be assigned to the literals $x$ such that $\forall y \varphi(x,y)$ is true.
	\item One seller $S^\psi_l$ for each $l =1,\ldots,L$. Each seller $S^\psi_l$ offers items $\psi_{lii'}$ for $i,i' = 1,\ldots,n$. The sellers correspond to the clauses of $\varphi(x,y)$ and below we describe how an allocation of the items from a seller of type $S^\psi$ corresponds to the truth value the corresponding clause evaluates to.
	\item One seller $S^{\gamma,\phi}$ who offers items $\gamma_j, \overline{\gamma}_j$ for $j = 1,\ldots,m$ as well as items $\phi_{li}$ for $l = 1,\ldots,L$ and $i = 1,\ldots,n$. The items of type $\gamma$ correspond to the possible values which can be assigned to literals $y$. The items $\phi_{li}$ are auxiliary items which indicate which clauses evaluate to false as a result of the assignment of $y$. While items of type $\psi$ already correspond to the truth assignments of the clauses, these additional auxiliary items are necessary in the proof for stability reasons since seller $S^{\gamma,\phi}$ now also needs to be part of any blocking coalition involving items corresponding to the truth assignment of clauses.
	\item One seller $S^{\lambda}$ who offers items $\lambda^k_i$ and $\overline{\lambda}^k_i$ for $i = 1,\ldots,n$ and $k=1,2$. These serve as auxiliary items to increase competition for buyers in order to drive up prices and deplete the budgets of buyers, as we will describe below	
\end{itemize}

We introduce the following short notations for bundles of items:
\begin{itemize}
	\item $\mathcal{T}^{\psi}_{l} = \left\{\psi_{li'i} | i,i' = 1,\ldots,n\right\}$
	\item $\mathcal{T}^{\psi,\phi}_{li} = \mathcal{T}^{\psi}_{l} \cup \left\{\phi_{li}\right\}$
	\item $\mathcal{F}^{\psi}_{li} = \left\{\psi_{lii'} | i' = 1,\ldots,n\right\}$s
	\item $\mathcal{F}^{\psi,\phi}_{l} = \mathcal{T}^{\psi,\phi}_{l} \cup \left\{\phi_{li} | i = 1,\ldots,n\right\}$
\end{itemize}

\begin{figure}
\begin{center}
\caption{Illustration of buyers' interests, only concerning items of type $\psi$}
\label{fig:complex_bundles}
\includegraphics[width=0.5\columnwidth]{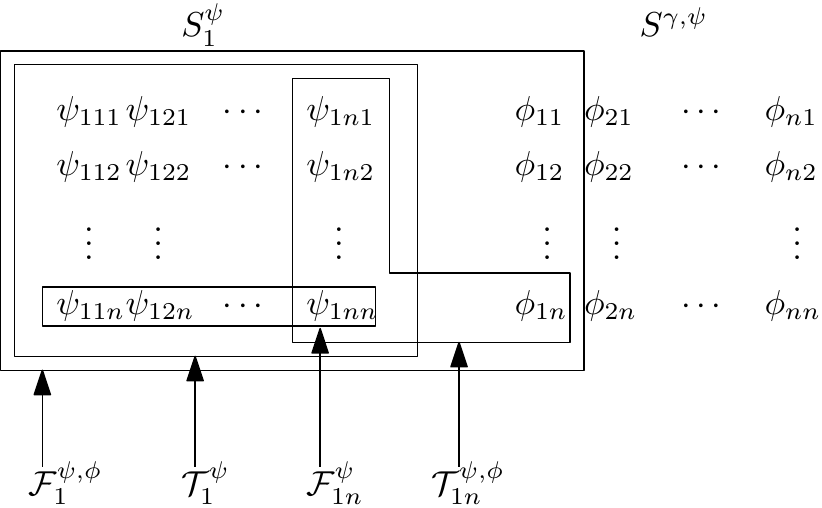}
\end{center}
\end{figure}

Figure \ref{fig:complex_bundles} illustrates an example for these bundles of items sold by $S^\psi_1$ and $S^{\gamma,\phi}$. It can be seen that the bundles intersect with each other in such a way, that if for any $i \in \left\{1,\ldots,n\right\}$, a bundle $\mathcal{F}^{\psi}_{li}$ is purchased by a buyer, no bundle $\mathcal{T}^{\psi,\phi}_{li'}$ can be purchased for any $i' \in \left\{1,\ldots,n\right\}$ and vice versa. Similarly, bundles $\mathcal{F}^{\psi,\phi}_l$ and $\mathcal{T}_l$ intersect with all other bundles.

Next, we define the buyers with their preferences and budgets. Let $T < \frac{1}{n}$, $U > nL$, $V > 4U$ and $W > 7n V$. First, we define buyers of type $B^{\mathcal{K}}$ and type $B^{\mathcal{M}}$ whose assignments will directly correspond to the logical values of the literals $x$
\begin{itemize}
	\item For $i = 1,\ldots,n$ let $B^{\mathcal{K}}_i$ be a buyer with a budget of $V + T$ and a value of $W$ for each of the following bundles:
		\begin{itemize}
			\item For $l = 1,\ldots,L$, bundle $\mathcal{K}_l := \left\{\chi_i\right\} \cup \mathcal{T}^{\psi,\phi}_{li}$
			\item For $l = 1,\ldots,L$, bundle $\overline{\mathcal{K}}_l := \left\{\overline{\chi}_i\right\} \cup \mathcal{T}^{\psi,\phi}_{li}$
		\end{itemize}
			He is interested in obtaining exactly one of these bundles and his value for obtaining one or more of the bundles is equal to the maximal value of his obtained bundles.
	\item For $i = 1,\ldots,n$ let $B^{\mathcal{M}}_i$ be a buyer with a budget of $2 V$ and a value of $2 V$ for the bundles
	\begin{itemize}
		\item $\mathcal{M}_i := \left\{\chi_i,\lambda^1_i,\lambda^2_i\right\} \cup \bigcup_{x_i \in C_l} \mathcal{F}^{\psi}_{li}$
		\item $\overline{\mathcal{M}}_i \left\{\overline{\chi}_i,\overline{\lambda}^1_i,\overline{\lambda}^2_i\right\} \cup \bigcup_{\overline{x}_i \in C_l} \mathcal{F}^{\psi}_{li}$
	\end{itemize}
	He is interested in exactly one of these bundles.
\end{itemize}

Buyers $B^{\mathcal{K}}$ and $B^{\mathcal{M}}$ are designed in such a way that for all $i \in \left\{1,\ldots,n\right\}$, buyer $B^{\mathcal{K}}_i$ will obtain one of the items $\left\{\chi_i,\overline{\chi}_i\right\}$, while buyer $B^{\mathcal{M}}_i$ obtains the other item. Whenever $B^{\mathcal{K}}_i$ buys $\chi_i$, this corresponds to an assignment of 'true' to the corresponding $x_i$ and whenever $B^{\mathcal{K}}_i$ buys $\overline{\chi}_i$ it corresponds to an assignment of 'false'. 
Buyers $B^{\mathcal{M}}_i$ obtains the opposite item (corresponding to its negation) as well as the bundles $\mathcal{F}^{\psi}_{li}$ for all $l$ which evaluate to 'false' due to the assignment of $B^{\mathcal{M}}_i$. The budgets and valuations are chosen in such a way, that non of the buyers described below can outbid buyers of type $B^{\mathcal{M}}_i$ at seller $S^\psi_l$, i.e. no bundles containing any item of $\mathcal{T}_l$ can be sold when $B^{\mathcal{M}}_i$ desires $\mathcal{F}^{\psi}_{li}$ for some $i \in \left\{1,\ldots,n\right\}$. In the following, we say that $B^{\mathcal{M}}_i$ {\it blocks} the bundle $\mathcal{T}_l$ (i.e. bundles $\mathcal{T}^{\psi,\phi}_{li}$ for all $i$ as well as bundle $\mathcal{F}^{\psi,\phi}_{l}$). We will see in the proof that when buyers of types $B^{\mathcal{G}}$ and $B^{\mathcal{K}}$ can only compete for unblocked bundles.

Additionally, we introduce the following auxiliary bidders who drive up prices in order to deplete the budgets of buyers $B^{\mathcal{K}}$ and $B^{\mathcal{M}}$.
\begin{itemize}
	\item For $i = 1,\ldots,n$, identical buyers $B^{\chi,1}_i$ and $B^{\chi,2}_i$ who are interested in one of $\chi_i$ or $\overline{\chi}_i$, have a valuation of $V$ for both, as well as a budget of $V$
	\item For $i = 1,\ldots,n$ one buyer $B^{\lambda,1}_i$ who has a budget of $U$ and a value of $V$ for bundle $\lambda^1_i$ and a value of $V-L$ for $\overline{\lambda}^1_i$
	\item For $i = 1,\ldots,n$ one buyer $B^{\lambda,2}_i$ who has a budget of $U$ and a value of $V-L$ for bundle $\lambda^2_i$ and a value of $V$ for $\overline{\lambda}^2_i$
\end{itemize}

The reason for including these auxiliary buyers and items is to bind an amount of $V$ of the budget of buyer $B^{\mathcal{K}}_i$ to purchase items from seller $S^{\lambda}$ such that he only has a budget of $T$ left to purchase his remaining items from sellers of type $S^\psi$ and from seller $S^{\gamma,\phi}$. In the following, we define the final set of buyers which compete with buyers $B^{\mathcal{M}}$ for these items.

\begin{itemize}
	\item For $j = 1,\ldots,m$ one buyer $B^{\mathcal{G}}_j$ who has a budget of $L$ and a value of 1 for each bundle:
		\begin{itemize}
			\item $\mathcal{G}_{jl} = \left\{\gamma_j\right\} \cup \mathcal{F}^{\psi,\phi}_{l}$ for each $l=1,\ldots,L$ with $\overline{Y}_j \in C_l$
			\item $\overline{\mathcal{G}}_{lj} = \left\{\overline{\gamma}_j\right\} \cup \mathcal{F}^{\psi,\phi}_{l}$ for each $l=1,\ldots,L$ with $Y_j \in C_l$
		\end{itemize} 
		His valuation for obtaining a larger bundle $G$ containing one ore more of the bundles defined above is equal to $\frac{\max\left\{|\mathcal{G}_l|,|\overline{\mathcal{G}}_l|\right\} - 1}{n(n+1)}$, i.e. the maximum number of bundles of type $\mathcal{G}$ and type $\overline{\mathcal{G}}$ he obtains. Thus, each buyer $B^{\mathcal{G}}_j$ is only interested in obtaining bundles which do not include both items, $\gamma_j$ and $\overline{\gamma}_j$. We refer to a bundle which includes only one of these items with a valuation of $k$ as a \textit{clean bundle of size $k$}. 
\end{itemize}

\begin{table}
	\centering
		\begin{tabular}{l|l|l}
  Bidder type & Value & Budget \\ \hline
  $B^{\mathcal{K}}$ & $W$ & $V+T$ \\
	$B^{\mathcal{M}}$ & $2V$ & $2V$ \\
	$B^{\mathcal{G}}$ & 1 & $L$ \\
  $B^{\chi,1}, B^{\chi,2}$ & $V$ & $V$ \\
  $B^{\lambda,1}$ & $V$ for $\lambda^1$ and $V-L$ for $\overline{\lambda}^1$ & $U$ \\
	$B^{\lambda,2}$ & $V-L$ for $\lambda^1$ and $V$ for $\overline{\lambda}^1$ &$U$ 
		\end{tabular}
	\caption{Values and budgets of key buyer types.}
	\label{tab:tbd}
\end{table}

These buyers are designed in a way such that they compete for all bundles $\mathcal{F}^{\psi,\phi}_l$ for which no buyer of type $B^{\mathcal{M}}_i$ wants to buy bundle $\mathcal{F}^{\psi}_{li}$ since the latter has a larger budget and higher valuation and thus can always outbid buyers of type $B^{\mathcal{G}}$. Whenever a buyer of type $B^{\mathcal{G}}$ obtains such a bundle, it corresponds to the corresponding clause to evaluate to 'false'. Buyers of type $B^{\mathcal{G}}$ maximize their welfare by purchasing as many of these packages as possible, corresponding to causing as many clauses to evaluate to 'false' as possible which are not do not already evaluate to 'false' due to buyers of type $B^{\mathcal{M}}$. Only if buyers of type $B^{\mathcal{G}}$ can not buy all of these bundles, buyers of type $B^{\mathcal{K}}$ can be assigned their bundles. Similarly to above, we say that a buyer of type $B^{\mathcal{G}}$ blocks bundle $\mathcal{T}^{\psi,\phi}_{l}$ for buyers of type $B^{\mathcal{K}}$ if he purchases a bundle which contains $\mathcal{F}^{\psi,\phi}_{l}$.

\subsection{Reduction}

In the following, we prove that there exists a core solution in $\textrm{CEx}_{\varphi(x,y)}$ with a social welfare of at least $n W$ if and only if $\exists x \forall y \varphi(x,y)$ is true. We refer to such a core solution as an $n W$-equilibrium. 

First, we will prove these auxiliary results:

\begin{enumerate}
	\item[] Lemma \ref{the:l1}: In an $n W\textrm{-equilibrium}$, for each $i = 1,\ldots,n$, buyer $B^{\mathcal{K}}_i$ obtains one of the bundles he values at $W$.
	\item[] Lemma \ref{the:l2}: In an $n W\textrm{-equilibrium}$, for each $i = 1,\ldots,n$, buyer $B^{\mathcal{K}}_i$ obtains one of the items $\chi^1_i$ or $\overline{\chi}^1_i$, buyer $B^{\mathcal{M}}_i$ obtains the complementary item and both pay $V$ to $S^\chi_i$.
	\item[] Lemma \ref{the:l3}: In an $n W\textrm{-equilibrium}$, for $i = 1,\ldots,n$ buyer $B^{\mathcal{M}}_i$ obtains all his required items from sellers $S^\psi$ and $S^{\lambda}$.
	\item[] Lemma \ref{the:l4}: In any core allocation, buyers of type $B^{\mathcal{G}}$ maximize the combined size of their clean bundles among the ones not blocked by buyers of type $B^{\mathcal{M}}$.
	\item[] Lemma \ref{the:l5}: There is a $n W\textrm{-equilibrium}$ if and only if for $j=1,\ldots,m$, buyers $B^{\mathcal{G}}_j$ are not able to block all the remaining bundles for buyers of type $B^{\mathcal{K}}$.
\end{enumerate}

Using these auxiliary results, we will be able to prove the main result.

\begin{lemma}
\label{the:l1}
In an $n W\textrm{-equilibrium}$, for each $i = 1,\ldots,n$, buyer $B^{\mathcal{K}}_i$ obtains one of the bundles he values at $W$.
\end{lemma}
\proof{Proof:}
Assume that a buyer $B^{\mathcal{K}}_i$ does not obtain his preferred bundle (and thus, the total welfare generated by the other buyers $B^{\mathcal{K}}_i$ for $i' \neq i$ is at most $(n-1) V$). Then, there is no way to achieve a social welfare of at least $n V$ since 
\[W > 7n V > \underbrace{2n V}_{\textrm{Buyers} B^{\mathcal{M}}} + \underbrace{2n V}_{\textrm{Buyers} B^{\chi}} + \underbrace{2n V}_{\textrm{Buyers} B^{\lambda}} + \underbrace{nL}_{\textrm{Buyers} B^{\mathcal{G}}}\]
which is an upper bound on the welfare achievable by all other buyers.  
\qed \endproof

Thus, in an $n W\textrm{-equilibrium}$, all buyers of type $B^{\mathcal{K}}$ obtain one of their desired bundles. As we described in the transformation, this is only possible, if there exists at least one $l \in \left\{1,\ldots,L\right\}$, for which neither buyers of type $B^{\mathcal{M}}$ nor of type $B^{\mathcal{G}}$ block the bundle $\mathcal{T}^{\psi,\phi}_{l}$.

The following lemma is a simple observation how auxiliary buyers of type $B^{\chi}$ are used to deplete the budget of buyers of type $B^{\mathcal{K}}$.

\begin{lemma}
\label{the:l2}
In an $n W\textrm{-equilibrium}$, for each $i = 1,\ldots,n$, buyer $B^{\mathcal{K}}_i$ obtains one of the items $\chi^1_i$ or $\overline{\chi}^1_i$, buyer $B^{\mathcal{M}}_i$ obtains the other item and both pay $V$ to $S^\chi_i$.
\end{lemma}
\proof{Proof:}
If either $B^{\mathcal{K}}_i$ or $B^{\mathcal{M}}_i$ would pay less than $V$, then either $B^{\chi,1}_i$ or $B^{\chi,2}_i$ could outbid them and obtain the respective items: In this case, seller $S^\chi_i$, all buyers which obtain items from $S^\chi_i$ (and in consequence all further buyers and sellers) can form a coalition and share the additional payment of the buyer of type $B^{\chi}$ such that all members of this coalitions improve their payoffs. Thus, an assignment where $B^{\mathcal{K}}_i$ obtains an item from $S^\chi_i$ but pays less than $V$ can not be in the core. Since all buyers of type $B^{\mathcal{K}}$ need to obtain one of these items in order to reach an $n W\textrm{-equilibrium}$, the Lemma holds.
\qed \endproof

\begin{lemma}
\label{the:l3}
In an $n W\textrm{-equilibrium}$, for $i = 1,\ldots,n$ buyer $B^{\mathcal{M}}_i$ obtains all his required items from sellers $S^\psi$ and $S^{\lambda}$.
\end{lemma}
\proof{Proof:}
Because of Lemma \ref{the:l2}, in an $n W\textrm{-equilibrium}$, $B^{\mathcal{M}}_i$ needs to pay $V$ for the item he obtains from seller $S^\chi_i$. Then, he has a budget of $V$ left to obtain the missing items from seller $S^{\lambda}$ and sellers $S^\psi_l$ in order to complete his desired bundle. He needs to purchase items of the form $\mathcal{F}^{\psi}_{il}$ from $S^\psi_l$ as well as either $\left\{\lambda^1_i,\lambda^2_i\right\}$ or $\left\{\overline{\lambda}^1_i,\overline{\lambda}^2_i\right\}$.  No other buyer $B^{\mathcal{M}}_j$ with $j \neq i$ is interested in obtaining any of these items since they appear in no bundles with positive valuation for them. The only buyers interested in a subset these items are buyers $B^{\mathcal{K}}_{i'}$ for $i' = 1,\ldots,n$ (who only have a budget of $T$ left due to Lemma \ref{the:l2}), buyers of type $B^{\mathcal{G}}$ (who have a budget of at most $L$ each) and buyers $B^{\lambda,1}_i$ and $B^{\lambda,2}_i$ (with a budget of $U$ each). Since
\[
	V > 4 U > \underbrace{T}_{\textrm{Buyers} B^{\mathcal{M}}} + \underbrace{nL}_{\textrm{Buyers} B^{\mathcal{G}}} + \underbrace{2 U}_{\textrm{Buyers} B^{\lambda}},
\]
buyer $B^{\mathcal{M}}_i$ can pay sellers $S^\psi$ and $S^{\lambda}$ enough to obtain his required items and there is no combination of buyers that can outbid $B^{\mathcal{M}}_i$ in order to form a coalition with the sellers such that all improve.
\qed \endproof

The previous Lemma \ref{the:l4} showed that for any $i =1,\ldots,n$, buyer $B^{\mathcal{M}}_i$ gets all the items he requires from sellers $S^\psi$ and $S^{\lambda}$ and in particular all his required bundles of the form $\mathcal{F}^{\psi}_{li}$. Thus, he blocks the bundle $\mathcal{T}_l$ and therefore also all bundles $\mathcal{F}^{\psi,\phi}_{l}$ for sellers of type $B^{\mathcal{G}}$.

\begin{lemma}
\label{the:l4}
In any core allocation, buyers of type $B^{\mathcal{G}}$ maximize the combined size of their clean bundles among the ones not blocked by buyers of type $B^{\mathcal{M}}$.
\end{lemma}

\proof{Proof:}
Assume that the maximum combined size of non-blocked clean bundles which can be obtained by buyers $B^{\mathcal{G}}$ is $K$ but that in the core solution buyers do only buy clean bundles with a combined size of $\kappa \leq K-1$. There are no other buyers except for those of type $B^{\mathcal{K}}$ which are interested in any of the items offered by sellers $S^{\lambda}$ or sellers $S^\psi_l$ for those $l$ for which $\mathcal{T}_l$ is not blocked. Since $1 > nT$, there can be a coalition of those sellers and buyers $B^{\mathcal{G}}$ which can generate a value of $K > \kappa$ and distribute the welfare such that all participants are better off. This is a contradiction to the allocation being in the core.
Then, if all buyers $B^{\mathcal{G}}$ pay the valuation of their obtained bundle to seller $S^{\lambda}$, there is no coalition among these sellers and buyers which want to deviate since $S^{\lambda}$ can never improve upon his payoff.
\qed \endproof

\begin{lemma}
\label{the:l5}
There is a $n V\textrm{-equilibrium}$ if and only if for $j=1,\ldots,m$, buyers $B^{\mathcal{G}}_j$ are not able to block all the remaining bundles for buyers of type $B^{\mathcal{K}}$.
\end{lemma}

\proof{Proof:}
For any $i,l$, buyer $B^{\mathcal{K}}_i$ is only able to obtain one of the sets $\mathcal{T}^{\psi,\phi}_{il}$ if it is neither blocked by a buyer $B^{\mathcal{M}}$ or $B^{\mathcal{G}}$. Thus, there is some $l$ for which all buyers can obtain these items if and only if buyers $B^{\mathcal{G}}$ do not block all of these bundles and as of Lemma \ref{the:l1} there is an $n W\textrm{-equilibrium}$ if and only if all buyers $B^{\mathcal{K}}$ obtain one of their bundles values at $W$
\qed \endproof

\begin{theorem}
\label{the:qsat2_noe}
There exists an $n W\textrm{-equilibrium}$ if and only if $\exists x \forall y \varphi(x,y)$ is true.
\end{theorem}
\proof{Proof:}
Consider an $n W\textrm{-equilibrium}$ and set $x_i$ to true if buyer $B^{\mathcal{K}}_i$ obtains item $\chi_i$ and set $x_i$ to false if he obtains $\overline{\chi}_i$. Then, buyer $B^{\mathcal{M}}_i$ obtains the negated item and bundles $\mathcal{F}^{\psi}_{li}$ for all clauses $C_l$ which evaluate to 'false' due to the assignment of $x_i$. This is equivalent to blocking the bundles $\mathcal{T}_l$ for buyers $B^{\mathcal{G}}$ who thus compete for the non-blocked bundles. Each combination of bundles obtained by buyer $B^{\mathcal{G}}_j$ resembles a number of clauses which can be made false by a truth assignment of $y_j$. If $B^{\mathcal{G}}_i$ obtains $\gamma_i$ this corresponds to an assignment of $y_i$ to true and if he obtains $\overline{\gamma}_i$ it corresponds to an assignment of $y_i$ to false. As of Lemma \ref{the:l3} and \ref{the:l4}, in any core allocation (and hence, especially in an $n W\textrm{-equilibrium}$ buyers $B^{\mathcal{G}}$ try to maximize their combined number of bundles not blocked by buyers of type $B^{\mathcal{M}}$ which is equivalent to blocking as many bundles as possible for buyers of type $B^{\mathcal{K}}$. This corresponds to assigning truth values to $y$ so that as many clauses as possible evaluate to false in $\varphi$. However, since by assumption the assignment results in an $n W\textrm{-equilibrium}$, buyers $B^{\mathcal{G}}$ are not successful in blocking all bundles because of Lemma $\ref{the:l5}$. Therefore, there is no assignment of variables $y$ such that $\varphi(x,y)$ can be set to false for this assignment of $x$.

Conversely, let $x$ be a truth assignment such that $\forall y \varphi(x,y)$ is true. Then, consider the following trades in the combinatorial exchange:
Trades for seller $S^\chi_i$ for $i = 1,\ldots,n$:
\begin{itemize}
	\item For $i = 1,\ldots,n$, if $x_i$ is true, assign to buyer $B^{\mathcal{K}}_i$ items $\chi_i$ for a price of $V$.
	\item For $i = 1,\ldots,n$, if $x_i$ is false, assign to buyer $B^{\mathcal{K}}_i$ items $\overline{\chi}_i$ for a price of $V$.
	\item For $i = 1,\ldots,n$, assign to buyer $B^{\mathcal{M}}_i$ the item not allocated to $B^{\mathcal{K}}_i$ for a price of $V$.
\end{itemize}

Trades for seller $S^{\lambda}$:
\begin{itemize}
	\item For $i = 1,\ldots,n$, if $x_i$ is true, assign to buyer $B^{\mathcal{M}}_i$ the items $\overline{\lambda}^1_i, \overline{\lambda}^2_i$ for a price of $2 U$, as well as item $\lambda^1_i$ to $B^{\lambda,1}_i$ and $\lambda^2_i$ to $B^{\lambda,2}_i$ for a price of $U$ each.
	\item For $i = 1,\ldots,n$, if $x_i$ is false, assign to buyer $B^{\mathcal{M}}_i$ the items $\lambda^1_i, \lambda^2_i$ for a price of $2 U$, as well as item $\overline{\lambda}^1_i$ to $B^{\lambda,1}_i$ and $\overline{\lambda}^2_i$ to $B^{\lambda,2}_i$ for a price of $U$ each.
	\end{itemize}

Further, assign to buyers $B^{\mathcal{M}}_i$ his remaining required items from sellers $S^\psi$, paying a price of $1$ to each seller he purchases from. Then, there is a $n W\textrm{-equilibrium}$ which extends these assignments. Similar to the first part of the proof, a buyer which blocks a bundle $\mathcal{T}_l$ for the other buyers corresponds to a truth assignment of the corresponding variable which results in the clause $l$ to evaluate to false. Since $\not\exists y: \neg \varphi(x,y)$ for the truth assignment of $x$, buyers $B^{\mathcal{M}}$ and $B^{\mathcal{G}}$ can not block all bundles for buyers $B^{\mathcal{K}}$, 
so each of them can obtain a bundle which he values at $W$. There is no coalition of buyers and sellers which want to deviate from this equilibrium: 
\begin{itemize}
	\item Buyers $B^{\mathcal{G}}$, sellers and $S^\psi$ and $S^{\gamma,\phi}$ can't form a coalition exclusively among themselves as of Lemma \ref{the:l5}.
	\item For all $i = 1,\ldots,n$, there exists no coalition including buyers $B^{\mathcal{M}}_i$ in which all participants can be made better off: Since $B^{\mathcal{M}}_i$ needs to pay sellers $S^\chi_i$ and $S^{\lambda}$ more money in order for them to join the coalition, he needs to pay less to the sellers $S^\psi$ he switches to. Those sellers are disjoint from the sellers $S^\psi$ he purchased from earlier. Thus, he can save at most $\frac{L}{2}$ units from switching which he needs to redistribute to $S^\chi_i$ and $S^{\lambda}$. However, since buyers $B^{\lambda,1}_i$ and $B^{\lambda,2}_i$ are affected by these trades as well, they need to be in the coalition as well and purchase items such that $S^{\lambda}$ can be made better off (since $\frac{L}{2} < 2D$, buyer $S^{\lambda}$ can not deviate only with $B^{\mathcal{M}}_i$). However, since for one of the two buyers, his new payoff is reduced by $L$, he will not agree to this coalition unless his payment is also reduced by at least $L$. However, since the second of these two buyers can not pay more as he is already capped by his budget, this is not possible.
	\item All other buyers and sellers can not deviate from the grand coalition on their own but need at least one buyer $B^{\mathcal{M}}$ in order for all members to achieve a higher payoff. As by the above, there is no coalition including a buyer $B^{\mathcal{M}}$ that can achieve this.
\end{itemize}
Thus, for a given truth assignment, there is a $n W\textrm{-equilibrium}$ and the proof is complete.
\qed \endproof
\end{APPENDICES}






\end{document}